\documentclass[twocolumn,preprintnumbers,prx]{revtex4}
\usepackage{graphicx}
\usepackage{dcolumn}
\usepackage{bm}
\usepackage{amsmath}
\usepackage{amssymb}
\setcitestyle{super}

\begin{document}

\title{Spin Casimir Effect in Non-collinear Quantum Antiferromagnets: Torque Equilibrium Spin Wave Approach}

\author{Z. Z. Du, H. M. Liu, Y. L. Xie, Q. H. Wang and J. -M. Liu}

\affiliation{Laboratory of Solid State Microstructures and Innovation Center of Advanced Microstructures, Nanjing University, Nanjing 210093, China}

\begin{abstract}
The Casimir effect is a general phenomenon in physics, which arises when the vacuum fluctuation of an arbitrary field is modified by static or slowly varying boundary.
However, its spin version is rarely addressed, mainly due to the fact that a macroscopic boundary  in quantum spin systems is hard to define.
In this article, we explore the spin Casimir effect induced by the zero-point fluctuation of spin waves in a general non-collinear ordered quantum antiferromagnet.
This spin Casimir effect results in a spin torque between local spins and further causes various singular and divergent results in the framework of spin-wave theory, which invalidate the standard $1/S$ expansion procedure.
Based on the spin Casimir torque interpretation, we develop a spin-wave expansion approach named as \emph{torque equilibrium spin wave theory} (TESWT). In this approach, the spin Casimir effect is treated in a self-consistent way, and the spin-wave expansion results are free from singularities and divergences.
A detailed spin-wave analysis of the antiferromagnetic spin-1/2 Heisenberg model on a spatially anisotropic triangular lattice is undertaken within our approach.
Our results indicate that the spiral order is only stable in the region $0.5<\alpha<1.2$, where $\alpha$ is the ratio of the coupling constants.
In addition, the instability in the region $1.2<\alpha<2$ is owing to the spin Casimir effect instead of the vanishing sublattice magnetization.
Furthermore, our method provides an efficient and convenient tool that can estimate the correct exchange parameters and outline the quantum phase diagrams, which can be useful for experimental fitting processes in frustrated quantum magnets.
\end{abstract}

\maketitle

\section{Introduction}
Low-dimensional quantum antiferromagnetic systems have witnessed a great deal of interest for a long time due to their deep connection with magnetic properties of high-temperature superconductors.~\onlinecite{RMP1,RMP2}
However, low-dimensional quantum spin systems are of interest in their own right as fruitful resources of novel and exotic quantum phases, such as valence bond solids~\onlinecite{Read1,Read2} and spin liquids (SL).~\onlinecite{RMP2,Balents1,Balents2,Wen,QM} The most indispensable ingredient in the emergence of these exotic quantum states is the quantum fluctuation caused by the $SU(2)$ commutation relation of the spin operators.
Frustration, on the other hand, acts as a very efficacious way of enhancing the quantum fluctuation effects and can even lead to the melting of magnetic long range order at zero temperature.~\onlinecite{Balents2,QM}
If such magnetic long range order survives, it is expected that the quantum fluctuation effects will be suppressed and can have only small influences on the thermodynamic properties of the system.~\onlinecite{QM}
Nevertheless, exceptions may occur, and as we shall see, quantum fluctuation can exhibit decisive consequences in non-collinear quantum antiferromagnets, despite the fact that the system is long range ordered.
This can occur because of the Casimir effect generated by the zero-point fluctuation in a non-collinear background, which may generate some emergent phenomena to be less touched so far.

The Casimir effect was originally discovered by Casimir in 1948, which states the presence of an attractive force between two parallel conducting plates placed in the vacuum.~\onlinecite{Ca1,Ca2}
This effect, which was described by Schwinger as one of the least intuitive consequences of quantum electrodynamics (QED), is actually ubiquitous in nature, covering many topics ranging from cosmology to condensed matter physics.~\onlinecite{CaRMP1,CaRMP2}
It arises when the quantum fluctuation of a general field (scalar, vector, spinor, or even tensor field) is modified by a static or slowly varying "boundary". This intriguing idea has generated continuing theoretical interest in generalized Casimir problems.~\onlinecite{CaFT1,CaFT2} The same type of Casimir effects have been predicted and discussed in many condensed matter systems such as quantum liquids~\onlinecite{Book1} and nanoparticle systems.~\onlinecite{CaRMP2}
The advantage of condensed matter systems as platforms to demonstrate the Casimir effects is the already known structure of the quantum vacuum, at least in principle.~\onlinecite{Book1} Moreover, various exotic quantum phases in condensed matter systems may allow different characteristics of the Casimir effects. In this respect, the low-dimensional quantum magnets seem to provide an ideal playground for dealing with the generalized Casimir problem and a spin version of this intriguing effect may be expected consequently.

The spin Casimir effect is the spin analog of the Casimir effect in vacuum, which describes various macroscopic Casimir force and torque that emergent from quantum spin systems.
Note the overall strength of the interaction generated by the Casimir Effect is proportional to the driving energy of quantum fluctuation ($\hbar$) and its scale is related to the correlation strength of the fluctuations.~\onlinecite{CaRMP1}
Thus, the Casimir force is expected to be strong and long-ranged in a system with strong fluctuation and long range correlation.
From this point of view, a system with highly degenerated ground states or in the vicinity of a quantum critical point is of particular interests and may generate rich Casimir physics.
Such as the instability of charge ordered states caused by spin Casimir effect in doped antiferromagnets, where the zero point spin-wave fluctuation induce a uniformly attractive force between hole clusters.~\onlinecite{CASC}
Another interesting example is the quantum fluctuation lifted massive classical degeneracy of the ground state, which is called "quantum order by disorder" (QObD).~\onlinecite{OBD1,OBD2}
In this case, the "boundary" is the long range ordered classical spin structure. The effective description of the Casimir effect is apparent in some cases. For example, the Casimir (QObD) effect  in a bilayer square-lattice Heisenberg antiferromagnetic model can be efficiently described by adding an additional term $(\textbf{S}_{i}\cdot \textbf{S}_{j})^{2}$ to the original Hamiltonian.~\onlinecite{Book2}
However, there are other cases where the detailed form of the Casimir physics is obscure and can be seen only by loop expansions.

In this article, we explore the emergence of spin Casimir effect in non-collinear ordered quantum magnets.
The appearance of this effect is due to the zero-point fluctuation in a non-collinear ordered spin structure and leads to the difference of measured ordering vector from the classical one. Although this difference has been discussed by several authors, its Casimir nature and related consequences have not yet been thoroughly investigated.~\onlinecite{Chubukov1,Da} This is mainly due to the fact that the spin Casimir effect is of order $O(1/S)$, which makes the modification of the ordering vector much smaller than the classical value. Thus it is usually negligible in classical ordered systems. In contrast, we predict in this paper that in some circumstances a standard spin-wave theory becomes no longer applicable due to the presence of the spin Casimir effect, even though the system is long-range ordered. In this sense, the spin Casimir effect is no longer negligible.
Furthermore, this effect can cause the spiral state instability, which is essentially different from other long-range order "melting" cases.
We consider a two-dimensional spatially anisotropic triangular spin-1/2 antiferromagnet for the sake of general interest and perform a concrete and well-controlled calculation. We believe that our results are equally applicable to other non-collinear ordered quantum systems with arbitrary spin value.

It is known that an isotropic triangular lattice Heisenberg antiferromagnet even for $S=1/2$ may order into the so-called $120^{\circ}$ state.~\onlinecite{QM,TL1,TL2,Chubukov2,Sasha1} As the spatially anisotropic exchange interaction is turned on, the spin Casimir torque emerges, imposing modification to the classical ordering vector. Surprisingly, a careful $1/S$ expansion in the anisotropic case shows that an usual perturbative estimation of the modification of the ordering vector becomes divergent near the quantum critical point and the one-loop expansions of the energy spectrum and sublattice magnetization are strongly singular.
These singular behaviors are believed to be the outcomes of the spin Casimir torque, and their appearance does not represent the onset of quantum disordered phases.
To fix this point, we develop a self-consistent approach in the framework of the spin-wave theory, giving correct ordering vector modification and excluding the singularities of the $1/S$ expansions. Based on this self-consistent approach, a quantum phase diagram is obtained, which is qualitatively consistent with previous numerical works.~\onlinecite{Zheng1,FRG,MSW1,MSW2}
More than that, detailed results can be obtained with our approach by calculation that is no harder than a linear spin-wave expansion.
Accordingly, our method can be considered as an efficient and fast experimental fitting tool for spiral phases.

The remainder of this article is organized as follows. In Section II we introduce the anisotropic triangular lattice antiferromagnetic model, which is simple but instructive and grasps the core ingredients of physics. Section III provides a brief review of the standard large-S expansion procedure and the formal definition of  the spin Casimir torque which describes the modification of the classical spin structure due to quantum fluctuation. The first-order $O(1/S)$ quantum correction for the spin-wave spectrum and its singular behavior are considered in Section IV A, while Section IV B is devoted to the calculation of the sublattice magnetization $M$ to the order of $O(1/S^{2})$, which is divergent due to the presence of the spin Casimir torque. In Section V we develope the torque equilibrium spin wave theory (TESWT), which is free of the aforementioned singularities. Several physical properties are calculated within our approach and a quantum phase diagram is obtained. And we discuss our scheme as an experimental exchange parameter fitting tool in Section VI. Finally, we draw our conclusions and discussions in Section VII.

\begin{figure}
  \includegraphics[width=8cm]{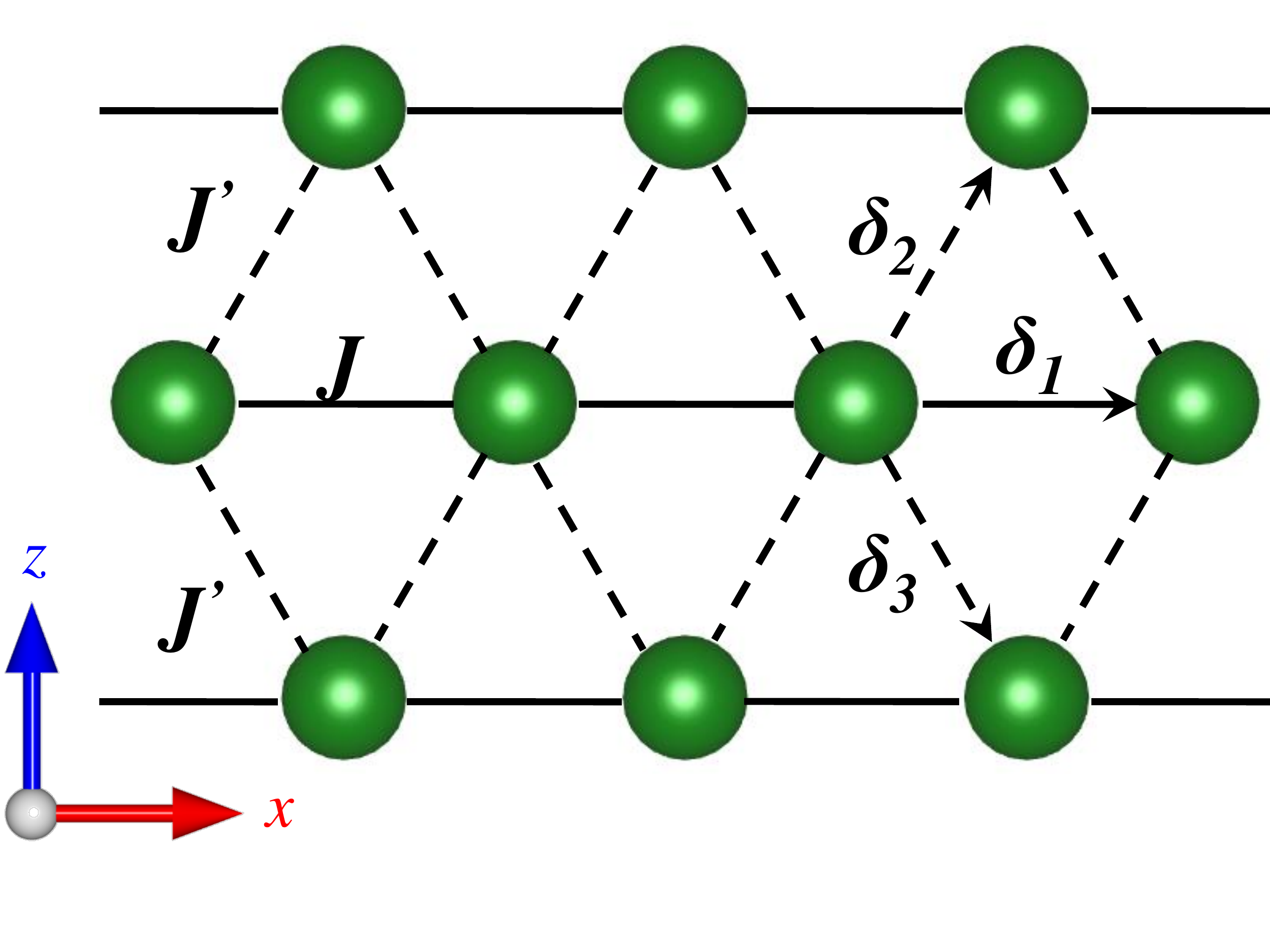}\\
  \caption{Exchange couplings between the different sites of the anisotropic triangular lattice.}
\end{figure}

\section{Anisotropic triangular lattice antiferromagnet}
The two-dimensional triangular lattice is the simplest realization of geometrical frustration where a spin-liquid has been suggested.
However, it is proved that the Heisenberg spins with the isotropic nearest-neighbor  antiferromagnetic interactions on such lattice display an long-range ordered state.~\onlinecite{QM,TL1,TL2} Nevertheless, the sublattice magnetization is highly reduced from its classical value due to the strong quantum fluctuation, indicating that small perturbations may destroy the long-range order and drive the system towards a quantum disordered state.~\onlinecite{Chubukov2,Sasha1} In this respect, different kinds of interactions have been studied on the triangular lattice for the potential realization of the spin liquid state.
Some of the most interested cases are the ring-exchange interaction,~\onlinecite{Ring1,Ring2} the next-nearest-neighbor interaction~\onlinecite{NN1,NN2,NN3,SBMF}, and the spatial anisotropic interaction.~\onlinecite{Zheng1,Zheng2,VMC,FRG,MSW1,MSW2,SBMF} The last case is particularly appreciated because of its applicability to real materials such as inorganic $Cs_{2}CuCl_{4}$~\onlinecite{CsCuCl1,CsCuCl2,CsCuCl3} and $Cs_{2}CuBr_{4}$~\onlinecite{CsCuBr1,CsCuBr2}, organic salts $\kappa-(BEDT-TTF)_{2}Cu_{2}(CN)_{3}$ and $\kappa-(BEDT-TTF)_{2}Cu_{2}[N(CN)]_{2}$.~\onlinecite{TL1,TL2}

The Heisenberg antiferromagnet on an anisotropic triangular lattice has its Hamiltonian:
\begin{equation}
    \hat{\mathcal{H}}=J\sum\limits^{\delta_{1}}_{\langle ij\rangle}\textbf{S}_{i}\cdot\textbf{S}_{j}+J^{\prime}\sum\limits^{\delta_{2},\delta_{3}}_{\langle ij\rangle}\textbf{S}_{i}\cdot\textbf{S}_{j}
\end{equation}
where $J$ is the interaction along the $\delta_{1}$, $J^{\prime}$ is the zig-zag interaction along the $\delta_{2,3}$ and the vectors $\delta_{i}$ connecting neighboring sites are shown in Fig.1. In this work, both $J$ and $J^{\prime}$ are positive and we denote the ratio of the coupling constants as $\alpha=J^{\prime}/J$. When $J=J^{\prime}$, the system is nothing but the isotropic triangular lattice Heisenberg antiferromagnetic model, whose ground state is a long-range ordered spiral state, i.e. the so called $120^{\circ}$ state. In the limit $J=0$, the system is equivalent to the isotropic square lattice Heisenberg antiferromagnetic model, whose ground state is also long-range ordered, the so-called Neel state. In the limit $J^{\prime}=0$, the system turns to the decoupled one-dimensional Heisenberg antiferromagnetic chains, where long-range order is forbidden even at zero temperature due to the Mermin-Wagner-Coleman theorem.~\onlinecite{MWC} In this case, the system is quantum disordered and shows many striking properties such as fractional excitation and power law correlation. Consequently, the related weakly coupled chain region ($J\gg J^{\prime}$) has attracted considerable attention.~\onlinecite{Balents2,TL1,TL2,Q1D1,Q1D2}

We first sketch the classical case. The classical ground state is usually simple and constitutes the foundation of the further spin-wave expansion. In the classical case, the quantum fluctuation is absent and spins are vectors rather than operators. In the whole parameter space the ground state is a general spiral structure whose magnetization $M_{i}$ on lattice point $r_{i}$ is given by:
\begin{equation}
    \hat{\textbf{m}}_{i}=\cos(\textbf{Q}_{cl}\cdot\textbf{r}_{i})\hat{\textbf{x}}+\sin(\textbf{Q}_{cl}\cdot\textbf{r}_{i})\hat{\textbf{z}}
\end{equation}
Here the spins are assumed to be in the $x-z$ plane and the classical ordering vector $\textbf{Q}_{cl}=(Q_{cl},0,0)$ is
\begin{eqnarray}
Q_{cl}=\Bigg\{\begin{array}{c}
         2\pi,~~~~~~~~~~~~~~~~~~~~~\alpha\geq2 \\
         \\
         \pi+2\arcsin(\alpha/2),~~\alpha<2
       \end{array}
\end{eqnarray}
which is determined by minimizing the classical ground state energy. This classical result is shown in Fig.2 and will be modified once the quantum fluctuation is considered.

\begin{figure}
  \includegraphics[width=8.5cm]{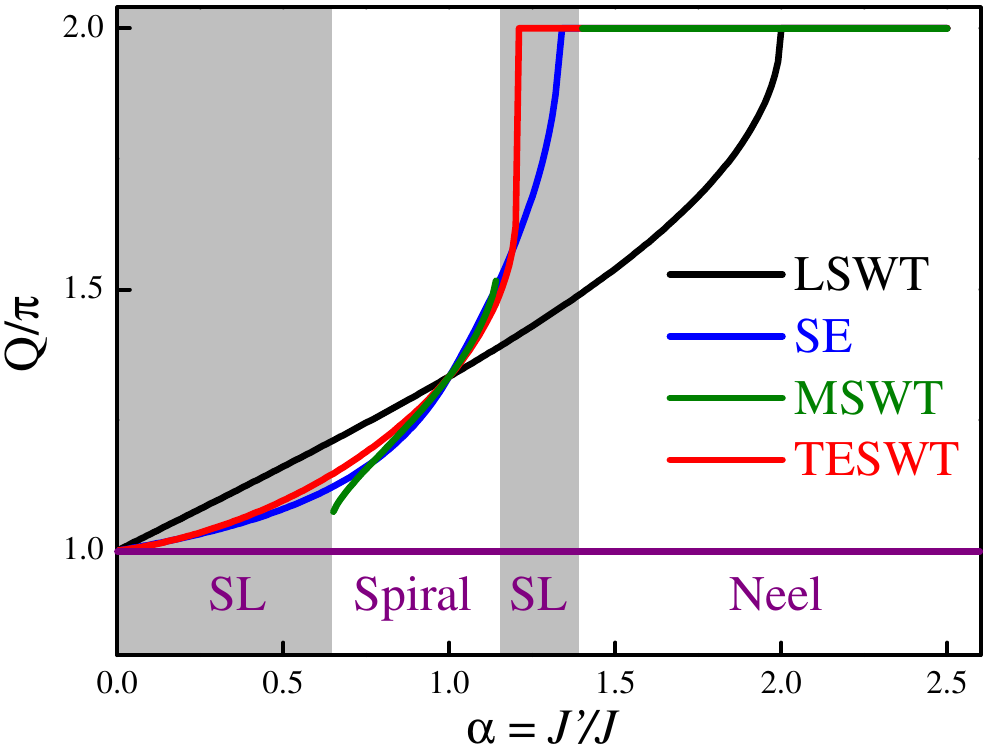}\\
  \caption{The ordering vector $Q$ (in units of $\pi$) of the optimal spiral state as a function of $\alpha=J^{\prime}/J$.
The black curve is the LSWT result, the blue curve is the SE results~\onlinecite{Zheng1}, the green curve is the MSWT results~\onlinecite{MSW1}, and the red curve is our torque equilibrium spin-wave theory (TESWT) result.
The gray regions denote the values of $\alpha$ where the modified spin-wave calculations fails to converge indicating the onset of the quantum disordered phases.~\onlinecite{MSW1}}
\end{figure}

Quite a number of theoretical approaches have been employed to treat the anisotropic triangular lattice Heisenberg model, such as linear spin-wave theory (LSWT),~\onlinecite{LSWT1,LSWT2} series expansions (SE),~\onlinecite{Zheng1,Zheng2} modified spin-wave theory (MSWT)~\onlinecite{MSW1} and density matrix renormalization group (DMRG).~\onlinecite{DMRG1,DMRG2} Besides, this model in the weakly coupled chain region has been studied by perturbative Bosonization and an effective Schrondinger equation approach.~\onlinecite{Q1D1,Q1D2} The core prediction in the weakly coupled chain region is the existence of the long-sought-after spin liquid phases, which seems even more elusive after the variational quantum Monte Carlo (VMC) method predicted two spin liquid phases in this region.~\onlinecite{VMC} Different from the low $\alpha$ region, it is well recognized that the QObD effect stabilizes considerably the Neel state over the classical model, moving the Neel phase from the classical region $\alpha\geq2$ to $\alpha\geq1.4$, although the phase boundary determination is technique-dependent. However, whether the quantum fluctuation spreads the transition point between the Neel and spiral phases into a spin liquid is still controversial.~\onlinecite{Zheng1,FRG,MSW1,MSW2} At first glance, this controversy has nothing to do with the so called spin Casimir effect which only appears in the spiral phase around $\alpha\approx1$. On the contrary, the emergence of the spin Casimir torque in the spiral state naturally explains the instability of the spiral state, which provide the foundation of the QDbO effect and the potential existence of the quantum disorder phase, to be discussed below.

\section{Large-S Expansion and Spin Casimir Torque}
The spin-wave approach starts from a classical spin configuration which minimizes the Heisenberg interaction and treats the quantum deviation from the ordered direction as collection of bosons.~\onlinecite{RMP1,Sasha1} In this work, the mapping from spin operators to bosons is performed via the Hermitian Holstein-Primakoff transformation in a twisted frame.~\onlinecite{HP} The resultant spin-wave Hamiltonian is given by
\begin{equation}
    \hat{\mathcal{H}}_{tot}=NE_{cl}+\hat{\mathcal{H}}_{sw} \nonumber
\end{equation}
with
\begin{equation}
    \hat{\mathcal{H}}_{sw}=\hat{\mathcal{H}}_{2}+\hat{\mathcal{H}}_{3}+\hat{\mathcal{H}}_{4}+O(S^{-1})
\end{equation}
where $E_{cl}=S^{2}J_{\textbf{Q}}$ is the classical ground state energy per spin, $\hat{H}_{n}$ denote the terms of order $S^{2-n/2}$ but the extension is only to the cubic and quartic anharmonic terms. Such an approximation is sufficient for calculation of the $O(1/S)$ order result of spin-wave spectrum and the $O(1/S^{2})$ order result of sublattice magnetization, which are our main interests.
In the Fourier transformed representation, the explicit expression of various terms in the Hamiltonian reads as
\begin{eqnarray}
    \hat{\mathcal{H}}_{2}&=&2S\sum\limits_{\textbf{k}}A_{\textbf{k}}a^{\dagger}_{\textbf{k}}a_{\textbf{k}}-\frac{B_{\textbf{k}}}{2}(a_{\textbf{k}}a_{-\textbf{k}}+a^{\dagger}_{\textbf{k}}a^{\dagger}_{-\textbf{k}}) \nonumber\\
    \hat{\mathcal{H}}_{3}&=&i\sqrt{2S}\sum\limits_{\textbf{k},\textbf{p}}\zeta_{\textbf{k}}(a^{\dagger}_{\textbf{k}+\textbf{p}}a_{\textbf{k}}a_{\textbf{p}}-a^{\dagger}_{\textbf{k}}a^{\dagger}_{\textbf{p}}a_{\textbf{k}+\textbf{p}}) \nonumber\\
    \hat{\mathcal{H}}_{4}&=&\frac{1}{4}\sum\limits_{\{\textbf{k}_{i}\}}\Big\{\big[(A_{\textbf{1}-\textbf{3}}+A_{\textbf{1}-\textbf{4}}+A_{\textbf{2}-\textbf{3}}+A_{\textbf{2}-\textbf{4}})  \nonumber\\
   &&-(B_{\textbf{1}-\textbf{3}}+B_{\textbf{1}-\textbf{4}}+B_{\textbf{2}-\textbf{3}}+B_{\textbf{2}-\textbf{4}})-(A_{\textbf{1}}+A_{\textbf{2}} \nonumber\\
   &&+A_{\textbf{3}}+A_{\textbf{4}})\big]a^{\dagger}_{\textbf{1}}a^{\dagger}_{\textbf{2}}a_{\textbf{3}}a_{\textbf{4}}\cdot\delta_{\textbf{1}+\textbf{2},\textbf{3}+\textbf{4}}+\frac{2}{3}(B_{\textbf{1}}+B_{\textbf{2}} \nonumber\\
   &&+B_{\textbf{3}})(a^{\dagger}_{\textbf{1}}a^{\dagger}_{\textbf{2}}a^{\dagger}_{\textbf{3}}a_{\textbf{4}}+a_{\textbf{1}}a_{\textbf{2}}a_{\textbf{3}}a^{\dagger}_{\textbf{4}})\cdot\delta_{\textbf{1}+\textbf{2}+\textbf{3},\textbf{4}}\Big\}
\end{eqnarray}
Here, $\textbf{1}, \textbf{2}...$ denote $\textbf{k}_{1}, \textbf{k}_{2}...$, and the following functions are introduced:
\begin{eqnarray}
    &&J_{\textbf{k}}=J\cos k_{x}+2J^{\prime}\cos \frac{k_{x}}{2}\cos \frac{\sqrt{3}}{2}k_{y}  \nonumber\\
    &&\eta_{\textbf{k}}=\frac{1}{2}(J_{\textbf{k}-\textbf{Q}}+J_{\textbf{k}+\textbf{Q}}),~~~~\zeta_{\textbf{k}}=\frac{1}{2}(J_{\textbf{k}-\textbf{Q}}-J_{\textbf{k}+\textbf{Q}})\nonumber\\
    &&A_{\textbf{k}}=\frac{1}{2}(J_{\textbf{k}}+\eta_{\textbf{k}}-2J_{\textbf{Q}}),~~B_{\textbf{k}}=\frac{1}{2}(J_{\textbf{k}}-\eta_{\textbf{k}})
\end{eqnarray}

On this basis, one may perform the $1/S$ perturbation expansion either following the formalism developed by Belyaev~\onlinecite{Da,Vei} or turning to the Bogoliubov¡¯s quasi-particle representation.~\onlinecite{Chubukov2,Sasha1} Here we follow the latter scheme, in which the quantum fluctuation induced singular behaviors are much more evident.

\begin{figure}
  \centering\includegraphics[width=7cm]{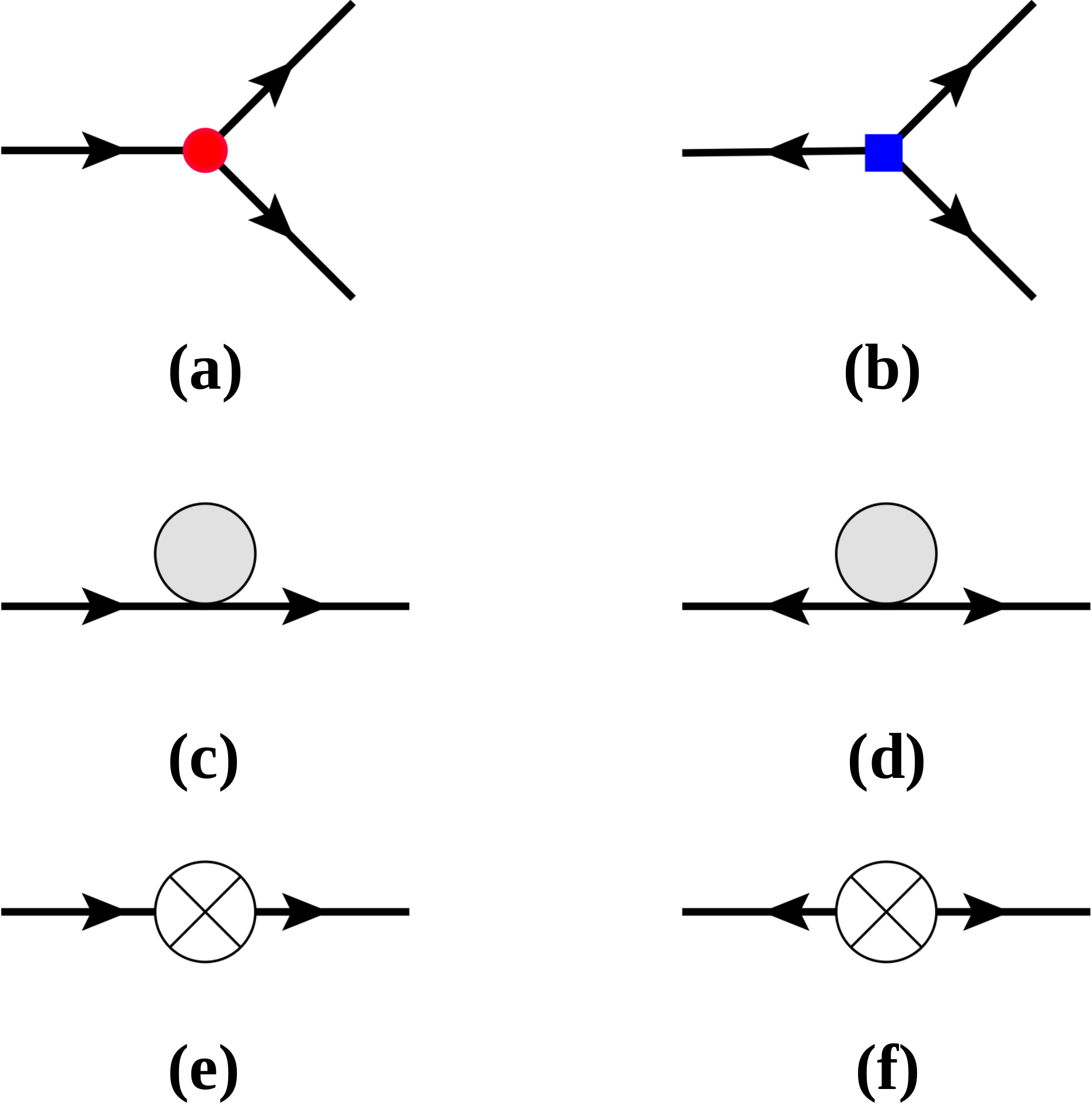}\\
  \caption{The lowest-order cubic vertices [(a) and (b)], Hartree-Fock vertices [(c) and (d)] and "counterterms" vertices [(e) and (f)].}
\end{figure}

The quasi-particle representation is related to the Holstein-Primakoff representation by a Bogolyubov transformation~\onlinecite{Wen,Book1,Book2}
\begin{equation}
    a_{\textbf{k}}=u_{\textbf{k}}b_{\textbf{k}}+v_{\textbf{k}}b^{\dagger}_{-\textbf{k}},~~~~
    a^{\dagger}_{\textbf{k}}=u_{\textbf{k}}b^{\dagger}_{\textbf{k}}+v_{\textbf{k}}b_{-\textbf{k}}
\end{equation}
under conditions $u^{2}_{\textbf{k}}-v^{2}_{\textbf{k}}=1$ and
\begin{equation}
    u^{2}_{\textbf{k}}+v^{2}_{\textbf{k}}=\frac{A_{\textbf{k}}}{\varepsilon_{\textbf{k}}},~~~~2u_{\textbf{k}}v_{\textbf{k}}=\frac{B_{\textbf{k}}}{\varepsilon_{\textbf{k}}}
\end{equation}
with
\begin{equation}
    \varepsilon_{\textbf{k}}=\sqrt{A^{2}_{\textbf{k}}-B^{2}_{\textbf{k}}}
\end{equation}
As a result, the linear spin-wave Hamiltonian takes the following diagonalized form:
\begin{equation}
   \hat{\mathcal{H}}_{2}=2S\sum\limits_{\textbf{k}}\varepsilon_{\textbf{k}}(b^{\dagger}_{\textbf{k}}b_{\textbf{k}}+\frac{1}{2})-\frac{A_{\textbf{k}}}{2}
\end{equation}
and the cubic term $\hat{\mathcal{H}}_{3}$ in the new representation is
\begin{eqnarray}
   \hat{\mathcal{H}}_{3}&=&i\sqrt{2S}\sum\limits_{\textbf{k},\textbf{p}}\Big[\frac{1}{2!}\Gamma_{1}(\textbf{p},\textbf{k}-\textbf{p};\textbf{k})b_{\textbf{k}}b^{\dagger}_{\textbf{k}-\textbf{p}}b^{\dagger}_{\textbf{p}} \nonumber\\
               &+&\frac{1}{3!}\Gamma_{2}(\textbf{p},-\textbf{k}-\textbf{p};\textbf{k})b^{\dagger}_{\textbf{p}}b^{\dagger}_{-\textbf{k}-\textbf{p}}b^{\dagger}_{\textbf{k}}-\textrm{H.c.}\Big]
\end{eqnarray}
The first term describes the magnon decay processes and is symmetric under permutation of two outgoing momenta.
The second term serves as a magnon source and is symmetric under permutation of all three outgoing momenta.~\onlinecite{Sasha1}
The explicit form of these two vertices are
\begin{eqnarray}
   \Gamma_{1}(\textbf{1},\textbf{2};\textbf{3})&=&\frac{-1}{2\xi}\Big[\zeta_{\textbf{1}}\kappa_{\textbf{1}}(\gamma_{\textbf{2}}\gamma_{\textbf{3}}+\kappa_{\textbf{2}}\kappa_{\textbf{3}})
                      +\zeta_{\textbf{2}}\kappa_{\textbf{2}}(\gamma_{\textbf{1}}\gamma_{\textbf{3}} \nonumber\\
                      &&+\kappa_{\textbf{1}}\kappa_{\textbf{3}})+\zeta_{\textbf{3}}\kappa_{\textbf{3}}(\gamma_{\textbf{1}}\gamma_{\textbf{2}}-\kappa_{\textbf{1}}\kappa_{\textbf{2}})\Big]
                      \nonumber\\
                       \nonumber\\
   \Gamma_{2}(\textbf{1},\textbf{2};\textbf{3})&=&\frac{1}{2\xi}\Big[\zeta_{\textbf{1}}\kappa_{\textbf{1}}(\gamma_{\textbf{2}}\gamma_{\textbf{3}}-\kappa_{\textbf{2}}\kappa_{\textbf{3}})
                     +\zeta_{\textbf{2}}\kappa_{\textbf{2}}(\gamma_{\textbf{1}}\gamma_{\textbf{3}}
                     \nonumber\\
                     &&-\kappa_{\textbf{1}}\kappa_{\textbf{3}})+\zeta_{\textbf{3}}\kappa_{\textbf{3}}(\gamma_{\textbf{1}}\gamma_{\textbf{2}}-\kappa_{\textbf{1}}\kappa_{\textbf{2}})\Big]
\end{eqnarray}
with
\begin{equation}
   \xi=\sqrt{\varepsilon_{\textbf{1}}\varepsilon_{\textbf{2}}\varepsilon_{\textbf{3}}},~~~
   \kappa_{\textbf{i}}=\sqrt{A_{\textbf{i}}+B_{\textbf{i}}},~~~
   \gamma_{\textbf{i}}=\sqrt{A_{\textbf{i}}-B_{\textbf{i}}}
\end{equation}
where $\textbf{i}\in(\textbf{1},\textbf{2},\textbf{3})$.

It is noted that the transformation of the quartic terms is rather cumbersome. Given that we are only interested in the one-loop results, the quartic terms can be conveniently decoupled using the Hartree-Fock approximation.~\onlinecite{Chubukov2,Sasha1,Da,Vei} Introducing the following Hartree-Fock averages in momentum space
\begin{equation}
    n_{\textbf{k}}=\langle a^{\dagger}_{\textbf{k}}a_{\textbf{k}}\rangle=\frac{A_{\textbf{k}}}{2\varepsilon_{\textbf{k}}}-\frac{1}{2},~~~~\Delta_{\textbf{k}}=\langle a_{\textbf{k}}a_{-\textbf{k}}\rangle=\frac{B_{\textbf{k}}}{2\varepsilon_{\textbf{k}}}
\end{equation}
the quartic terms turn to the form $\hat{\mathcal{H}}_{4}=E_{4}+\delta\hat{\mathcal{H}}_{2}$, where $E_{4}$ is the Hartee-Fock corrections to the ground state energy and $\delta\hat{\mathcal{H}}_{2}$ is the $1/S$ modification to the harmonic spin-wave Hamiltonian with the form
\begin{equation}
   \delta\hat{\mathcal{H}}_{2}=\sum\limits_{\textbf{k}}\delta\varepsilon_{\textbf{k}}b^{\dagger}_{\textbf{k}}b_{\textbf{k}}
                       -\frac{O_{\textbf{k}}}{2}(b_{\textbf{k}}b_{-\textbf{k}}+b^{\dagger}_{\textbf{k}}b^{\dagger}_{-\textbf{k}})
\end{equation}
where
\begin{eqnarray}
   \delta\varepsilon_{\textbf{k}}&=&(u^{2}_{\textbf{k}}+v^{2}_{\textbf{k}})\delta A_{\textbf{k}}-2u_{\textbf{k}}v_{\textbf{k}}\delta B_{\textbf{k}} \nonumber\\
    O_{\textbf{k}}&=&(u^{2}_{\textbf{k}}+v^{2}_{\textbf{k}})\delta B_{\textbf{k}}-2u_{\textbf{k}}v_{\textbf{k}}\delta A_{\textbf{k}}
\end{eqnarray}
and
\begin{eqnarray}
    \delta A_{\textbf{k}}&=&A_{\textbf{k}}+\sum\limits_{\textbf{p}}\frac{1}{\varepsilon_{\textbf{p}}}\Big[A_{\textbf{p}}(A_{\textbf{k}-\textbf{p}}-A_{\textbf{k}}-A_{\textbf{p}}-B_{\textbf{k}-\textbf{p}}) \nonumber\\
    &&+B_{\textbf{p}}(\frac{B_{\textbf{k}}}{2}+B_{\textbf{p}})\Big]\nonumber\\
    \delta B_{\textbf{k}}&=&B_{\textbf{k}}-\sum\limits_{\textbf{p}}\frac{1}{\varepsilon_{\textbf{p}}}\Big[B_{\textbf{p}}(A_{\textbf{k}-\textbf{p}}-\frac{A_{\textbf{k}}}{2}-A_{\textbf{p}}-B_{\textbf{k}-\textbf{p}}) \nonumber\\
    &&+A_{\textbf{p}}(B_{\textbf{k}}+B_{\textbf{p}})\Big]
\end{eqnarray}

The effective Hamiltonian that combines all the terms together now reads
\begin{eqnarray}
   \hat{\mathcal{H}}_{eff}&=&\sum\limits_{\textbf{k}}\Big[(2S\varepsilon_{\textbf{k}}+\delta\varepsilon_{\textbf{k}})b^{\dagger}_{\textbf{k}}b_{\textbf{k}}-\frac{O_{\textbf{k}}}{2}(b_{\textbf{k}}b_{-\textbf{k}}+b^{\dagger}_{\textbf{k}}b^{\dagger}_{-\textbf{k}})\Big]\nonumber\\
                &&+i\sqrt{2S}\sum\limits_{\textbf{k},\textbf{p}}\Big[\frac{1}{2!}\Gamma_{1}(\textbf{p},\textbf{k}-\textbf{p};\textbf{k})b_{\textbf{k}}b^{\dagger}_{\textbf{k}-\textbf{p}}b^{\dagger}_{\textbf{p}} \nonumber\\
               &+&\frac{1}{3!}\Gamma_{2}(\textbf{p},-\textbf{k}-\textbf{p};\textbf{k})b^{\dagger}_{\textbf{p}}b^{\dagger}_{-\textbf{k}-\textbf{p}}b^{\dagger}_{\textbf{k}}-\textrm{H.c.}\Big]
\end{eqnarray}
which provides the basis for the $1/S$ perturbative expansion. The related Feynman diagrams are shown in Fig.3(a)-(d).

At the same time,  the $1/S$ expansion contributes to the corrections of the ground state energy as well.
This modification comes from the zero-point fluctuation of magnon, which is the fluctuating vacuum of our Casimir problem.
For the sake of simplicity, we only consider the first order corrections to the vacuum energy per spin
\begin{equation}
   E_{vac}=E_{cl}+E_{2}=S^{2}J_{\textbf{Q}}+S(J_{\textbf{Q}}+\sum\limits_{\textbf{k}}\varepsilon_{\textbf{k}})
\end{equation}
Here $E_{2}$ is the energy correction from the harmonic spin-wave fluctuation.

With this vacuum energy modification, the ordering vector of the system should be determined by
minimizing the modified vacuum energy $E_{vac}$ via $\delta E_{vac}/\delta \textbf{Q}=0$.
However, this simple variational equation can not be solved directly due to the fact that the spin-wave spectrum function $\varepsilon_{k}$ is only well-defined at $\textbf{Q}=\textbf{Q}_{cl}$.
Thus, the variation is normally treated approximately as an expansion around $\textbf{Q}_{cl}$. The $1/S$ order result is
\begin{equation}
   \textbf{Q}=\textbf{Q}_{cl}+\textbf{Q}_{1}
\end{equation}
with
\begin{equation}
   \textbf{Q}_{1}=-\frac{1}{2S}\Bigg[\frac{\partial^{2}J_{\textbf{Q}}}{\partial \textbf{Q}^{2}}\Bigg]^{-1}
                \sum\limits_{\textbf{k}}\frac{A_{\textbf{k}}+B_{\textbf{k}}}{\varepsilon_{\textbf{k}}}\cdot\frac{\partial J_{\textbf{k}+\textbf{Q}}}{\partial \textbf{Q}}\Bigg|_{\textbf{Q}_{cl}}
\end{equation}
These ordering vectors have only the $x$ component in our case via $\textbf{Q}_{1}=(Q_{1},0,0)$ and $\textbf{Q}=(Q,0,0)$.
This result seems reasonable and is usually treated as the new ordering vector of the system.~\onlinecite{Chubukov1,Da}
However, this is not the case when the system is in the vicinity of a quantum critical point.
The first order correction $Q_{1}$ goes to infinity as $\alpha$ approaches the classical spiral/Neel critical point $\alpha=2$, which is shown in Fig.4.

\begin{figure}
  \centering\includegraphics[width=7cm]{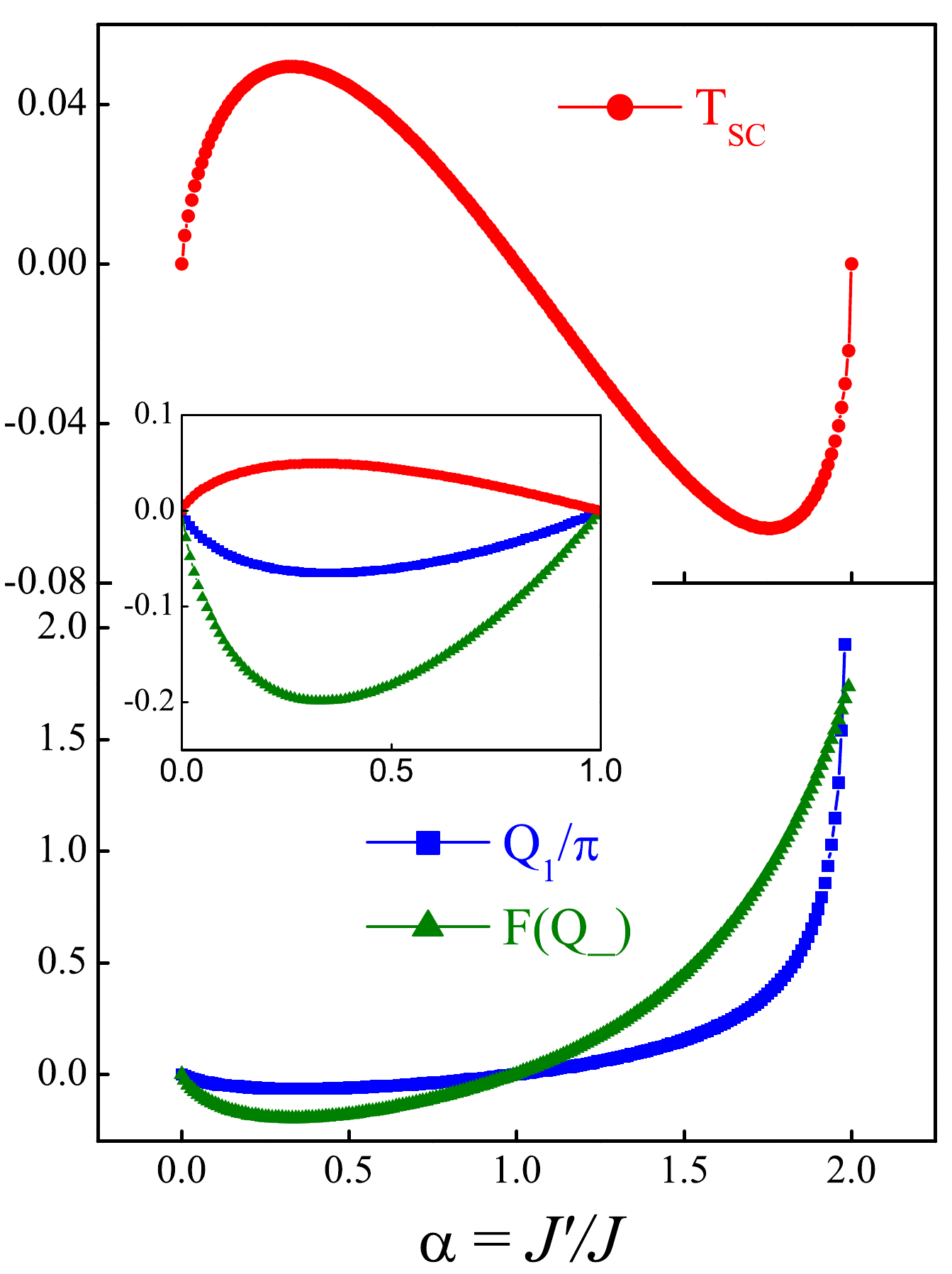}\\
  \caption{$T_{sc}$, $Q_{1}/\pi$ and $F(Q_{-})$ as a function of $\alpha$. Here $Q_{-}=Q-0^{+}$. The insert is these functions in region $0<\alpha<1$.
           Here both $Q_{1}/\pi$ and $F(Q_{-})$ are divergent when approaching $\alpha=2$ as described in the main text.}
\end{figure}

Clearly, this result is unphysical and needs to be regularized. Here we propose a torque description of the spin Casimir effect, which is analogous with the force description of the conventional Casimir problems in QED.~\onlinecite{CaRMP1,CaRMP2} To access a quantitative description, we define the spin Casimir torque as
\begin{equation}
   \textbf{T}_{sc}(\textbf{Q})=\sum\limits_{\textbf{k}}\Bigg\langle\Psi_{vac}\Bigg|\frac{\partial\hat{\mathcal{H}}_{sw}}{\partial \textbf{Q}}\Bigg|\Psi_{vac}\Bigg\rangle
\end{equation}
where $|\Psi_{vac}\rangle$ represents the quasi-particle vacuum state.
Notice that $\mathcal{T}_{sc}$ is a function of $\textbf{Q}$ defined on bonds and represents the tendency of the modification to the relative orientation of each spin.
One can have the same definition for a QObD system, which is equivalent with other conventional description methods.
The main advantage of the torque description is to provide a much more intuitive picture. As an example, the resultant $1/S$ order spin Casimir torque at $\textbf{Q}_{cl}$ is
\begin{equation}
   \textbf{T}_{sc}(\textbf{Q}_{cl})=\frac{S}{2}\sum\limits_{\textbf{k}}\frac{A_{\textbf{k}}+B_{\textbf{k}}}{\varepsilon_{\textbf{k}}}\cdot\frac{\partial J_{\textbf{k}+\textbf{Q}}}{\partial \textbf{Q}}\Bigg|_{\textbf{Q}_{cl}}
\end{equation}
And in our case $\textbf{T}_{sc}(\textbf{Q}_{cl})$ has only one component noted as $T_{sc}(Q_{cl})$.
Clearly, this spin Casimir torque is well-defined in the whole parameter space as shown in Fig.4.
And the zero ordering vector modification at $\alpha=1$ is easily explained by the zero spin Casimir torque due to the triangular symmetry.
For $\alpha\neq1$, the spin Casimir torque tends to arrange the spins connected by the strongest bonds collinearly.
This tendency is consistent with the conventional statement that fluctuation favors collinear spins.
Based on the torque interpretation and noting that $S^{2}\partial^{2}J_{\textbf{Q}}/\partial Q^{2}$ is just the classical spin stiffness $\rho_{s}$, the complex Eq.(21) is nothing but the Hooke's law for spin system
\begin{equation}
   T_{sc}(Q_{cl})=-\rho_{s}Q_{1}
\end{equation}
Additionally, the reason of the divergence of $Q_{1}$ in approaching the critical point ($\alpha=2$) is also apparent.
The transition at $\alpha=2$ is continuous and the spin-wave velocity for Goldstone excitation vanishes at the transition.
When the system approaches $\alpha=2$, the quantum fluctuation dominates and the spin stiffness quickly goes to zero.~\onlinecite{LSWT1,LSWT2}
Although $T_{sc}$ also goes to zero as $\alpha\rightarrow2$, its Casimir nature is enhanced as fluctuation dominates.
As a consequence, it approaches zero much more slowly than $\rho_{s}$, and eventually the resultant $Q_{1}$ blows up around the critical point, exhibiting the singularity at the point.
One may suggest that this singular behavior of $Q_{1}$ is understandable because the divergence of the fluctuations near a Lifshitz point is well known.~\onlinecite{Luca} As a consequence, the spin-wave theory should be invalid when near a quantum critical point.~\onlinecite{RMP1,QM}
However, the difficulty for the spin-wave theory is the existence of other singularities even far away from the critical point, in the presence of the spin Casimir torque.

\section{Spin Casimir effect induced Spin-wave Singularity}
The core ingredient of the spin-wave theory is the expansion in powers of $1/S$ around the classical saddle point.
Strictly, this theory is only well-defined in the large-S limit ($S\gg1$).~\onlinecite{RMP1,QM} and thus less effective for low-dimensional quantum spin systems, as quantum spin fluctuations typically increase in reduced space dimensions and for small spin quantum numbers $S$.
It is surprising to observe that the standard spin-wave approach can give very accurate description of the zero-temperature physics of a number of low-dimensional spin models such as the $S=1/2$ Heisenberg antiferromagnets on square and triangular lattices.~\onlinecite{RMP1,QM,Chubukov2,Sasha1} In this sense, this expansion approach can still be considered as a useful technique if an ordered state is observed. Nevertheless, it fails once the spin Casimir effect is taken into account, to be demonstrated here.

In the subsequent two subsections we will show that the one-loop expansions of the energy spectrum and sublattice magnetization are strongly singular and these singular behaviors are related to the spin Casimir torque. To access the explicit structure of these singular behaviors and show the breakdown of the conventional $1/S$ expansion procedure, we first ignore the spin Casimir effect and $\textbf{Q}$ is identified as $\textbf{Q}_{cl}$.

\subsection{Spin-wave Spectrum}
Perturbative expansion for the spin-wave spectrum has to be performed order by order in $1/S$ and takes into account all the quantum corrections of the same order.~\onlinecite{RMP1,QM}
Only in this manner one can ensure cancellation of all possible divergences in the individual contributions and preserve the Goldstone theorem.~\onlinecite{QFT} The first order correction is straightforward. The new pole of the magnon Green's function is determined by the so called Dyson equation~\onlinecite{Sasha1}
\begin{equation}
   \varepsilon=\varepsilon_{\textbf{k}}+\frac{1}{2S}\big[\delta\varepsilon_{\textbf{k}}+\Sigma^{a}_{3}(\textbf{k},\varepsilon)+\Sigma^{b}_{3}(\textbf{k},\varepsilon)\Big]
\end{equation}
with
\begin{eqnarray}
\Sigma^{a}_{3}(\textbf{k},\varepsilon)&=&\frac{1}{2}\sum_{\textbf{p}}\frac{|\Gamma_{1}(\textbf{p};\textbf{k})|^{2}}{\varepsilon-\varepsilon_{\textbf{p}}-\varepsilon_{\textbf{k}-\textbf{p}}+i0^{+}}  \nonumber\\
\Sigma^{b}_{3}(\textbf{k},\varepsilon)&=&-\frac{1}{2}\sum_{\textbf{p}}\frac{|\Gamma_{2}(\textbf{p};\textbf{k})|^{2}}{\varepsilon+\varepsilon_{\textbf{p}}+\varepsilon_{\textbf{k}+\textbf{p}}-i0^{+}}
\end{eqnarray}
The diagrammatic representations of the normal self-energies from the cubic terms are shown in Fig.5(a) and 5(b).

\begin{figure}
  \centering\includegraphics[width=8.5cm]{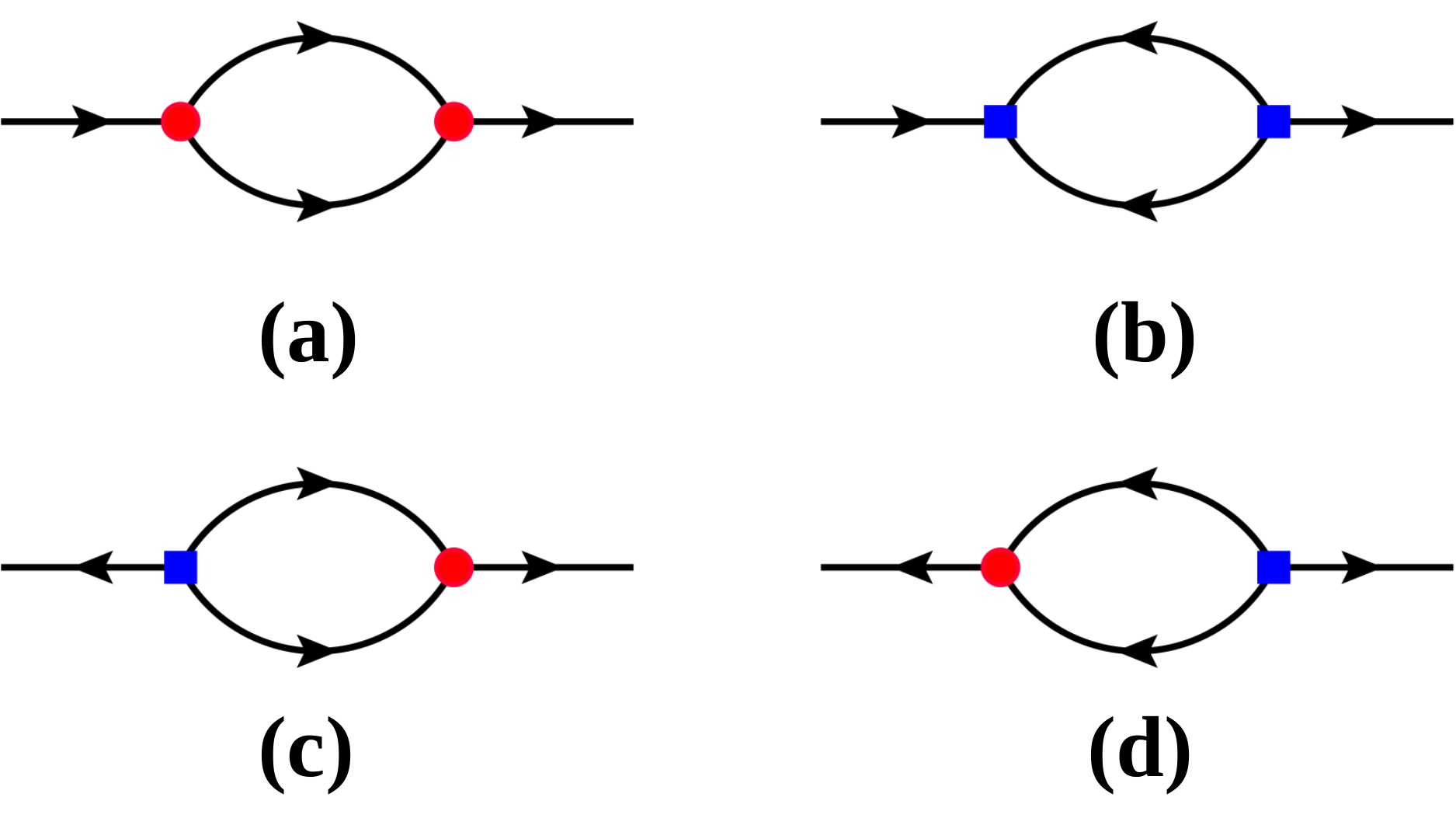}\\
  \caption{The lowest-order normal [(a) and (b)] and anomalous [(c) and (d)] magnon self-energies generated by cubic terms.}
\end{figure}

This equation can be solved either self-consistently through the off-shell approximation or by replacing $\varepsilon$ with linear spin-wave spectrum $\varepsilon_{\textbf{k}}$, i.e. the so called on-sell approximation. The $1/S$ order correction to the spectrum $F(\textbf{k})$ is obtained within the on-shell approximation, which leads to the following expression for the renormalized spectrum:
\begin{equation}
   F(\textbf{k})=\varepsilon_{\textbf{k}}+\frac{1}{2S}\Big[\delta\varepsilon_{\textbf{k}}+\Sigma^{a}_{3}(\textbf{k},\varepsilon_{\textbf{k}})+\Sigma^{b}_{3}(\textbf{k},\varepsilon_{\textbf{k}})\Big]
\end{equation}
Based on this expression, the one-loop spin-wave spectrum $F(\textbf{k})$ can be easily obtained by numerical integration of self-energies.
However, the numerical result shows the singularity of the spectrum at $\textbf{k}=\textbf{Q}$ (as shown in Fig.6.) and absence of the Goldstone mode, while the Goldstone mode is usually expected at every order of the perturbative expansion.~\onlinecite{QM,Chubukov2,Sasha1,QFT}
Thus the absence of the Goldstone mode and the instead appearance of the singular behavior are quite unusual.

To understand this singularity, a careful examination on all the contributions is needed. Regarding the Goldstone excitations, the terms proportional to $\varepsilon_{\textbf{k}}$ can be ignored safely. The resultant explicit form of the self-energies in the expression of $F(\textbf{k})$ is
\begin{eqnarray}
   \delta\varepsilon_{\textbf{k}}&\approx&\frac{1}{2\varepsilon_{\textbf{k}}}\Big[\kappa^{2}_{\textbf{k}}G_{0}(\textbf{k})+\gamma^{2}_{\textbf{k}}G_{Q}(\textbf{k})\Big] \nonumber\\
   \Sigma^{a}_{3}+\Sigma^{b}_{3}&\approx&-\frac{1}{4}\Big[\kappa^{2}_{\textbf{k}}L_{0}(\textbf{k})+\gamma^{2}_{\textbf{k}}L_{Q}(\textbf{k})\Big]
\end{eqnarray}
with
\begin{eqnarray}
   G_{0}(\textbf{k})&=&\sum_{\textbf{p}}\frac{\kappa^{2}_{\textbf{p}}\gamma^{2}_{\textbf{k}-\textbf{p}}}{\varepsilon_{\textbf{p}}}-\varepsilon_{\textbf{p}}   \nonumber\\
   G_{Q}(\textbf{k})&=&\sum_{\textbf{p}}\frac{\gamma^{2}_{\textbf{p}}\gamma^{2}_{\textbf{k}-\textbf{p}}}{\varepsilon_{\textbf{p}}}-\varepsilon_{\textbf{p}}   \nonumber\\
   L_{0}(\textbf{k})&=&\sum_{\textbf{p}}\frac{\Gamma_{0}^{2}}{\varepsilon_{\textbf{p}}+\varepsilon_{\textbf{k}-\textbf{p}}}\nonumber\\
   L_{Q}(\textbf{k})&=&\sum_{\textbf{p}}\frac{\Gamma_{Q}^{2}}
              {\varepsilon_{\textbf{p}}+\varepsilon_{\textbf{k}-\textbf{p}}}
\end{eqnarray}
and
\begin{eqnarray}
   &&\Gamma_{0}=\frac{1}{\xi}\Big[\zeta_{\textbf{k}}(\kappa_{\textbf{p}}\kappa_{\textbf{k}-\textbf{p}}-\gamma_{\textbf{p}}\gamma_{\textbf{k}-\textbf{p}})-(\zeta_{\textbf{p}}+\zeta_{\textbf{k}-\textbf{p}})\kappa_{\textbf{p}}\kappa_{\textbf{k}-\textbf{p}}\Big]
           \nonumber\\
   &&\Gamma_{Q}=\frac{1}
              {\xi}\big[\zeta_{\textbf{p}}\kappa_{\textbf{p}}\gamma_{\textbf{k}-\textbf{p}}+\zeta_{\textbf{k}-\textbf{p}}\gamma_{\textbf{p}}\kappa_{\textbf{k}-\textbf{p}}\big]
\end{eqnarray}
These equations allow a straightforward examination of the Goldstone mode at $\textbf{k}=\textbf{0}$ and $\textbf{k}=\textbf{Q}$.
Noting that $\gamma_{\textbf{0}}=0$ and $\zeta_{\textbf{0}}=0$, it is easy to prove $G_{0}(\textbf{0})/\varepsilon_{\textbf{0}}=0$ and $L_{0}(\textbf{0})=0$.
Hence, as $k\rightarrow0$ we have
\begin{equation}
   F(\textbf{0})=\frac{\kappa^{2}_{\textbf{0}}}{2S}\cdot\Bigg[\frac{G_{0}(\textbf{k})}{2\varepsilon_{\textbf{k}}}-\frac{L_{0}(\textbf{k})}{4}\Bigg]_{\textbf{k}\rightarrow\textbf{0}}=0
\end{equation}
Indeed, the Goldstone mode is preserved at $\textbf{k}=\textbf{0}$.

In contrast to the $\textbf{k}=\textbf{0}$ case, the Goldstone mode usually appears by cancelation among several terms.~\onlinecite{Chubukov2,Sasha1}
Notice that $\kappa_{\textbf{Q}}=0$ we obtain
\begin{equation}
   F(\textbf{Q})=\frac{\gamma^{2}_{\textbf{Q}}}{2S}\cdot\Bigg[\frac{G_{Q}(\textbf{k})}{2\varepsilon_{\textbf{k}}}-\frac{L_{Q}(\textbf{k})}{4}\Bigg]_{\textbf{k}\rightarrow \textbf{Q}}
\end{equation}
In this case, $G_{Q}(\textbf{Q})$ and $H_{Q}(\textbf{Q})$ do not equal zero in an obvious way, the resultant self-energies are divergent at $\texttt{k}=\textbf{Q}$.
To get the exact analytic structure, we introduce the following useful equalities
\begin{eqnarray}
  \zeta_{\textbf{k}}&=&\kappa^{2}_{\textbf{k}-\textbf{Q}}-\gamma^{2}_{\textbf{k}}  \nonumber\\
  \zeta_{\textbf{p}}&=&\kappa^{2}_{\textbf{k}-\textbf{p}}-\gamma^{2}_{\textbf{p}}+\delta_{1} \nonumber\\
  \zeta_{\textbf{k}-\textbf{p}}&=&\kappa^{2}_{\textbf{p}}-\gamma^{2}_{\textbf{k}-\textbf{p}}+\delta_{2}  \nonumber\\
  \delta_{1}&=&\kappa^{2}_{\textbf{p}-\textbf{Q}}-\kappa^{2}_{\textbf{k}-\textbf{p}} \nonumber\\
  \delta_{2}&=&\kappa^{2}_{\textbf{k}-\textbf{p}-\textbf{Q}}-\kappa^{2}_{\textbf{p}}
\end{eqnarray}
noting that $\delta_{1}$ and $\delta_{2}$ equal zero at $\textbf{k}=\textbf{Q}$. The expansion of $F(\textbf{Q})$ is straightforward with the aid of these equalities, and the final result is
\begin{eqnarray}
    F(\textbf{Q})&=&\frac{A_{\textbf{Q}}}{2S}\sum_{\textbf{p}}\frac{A_{\textbf{p}}+B_{\textbf{p}}}{\varepsilon_{\textbf{p}}}\frac{\partial J_{\textbf{p}+\textbf{Q}}}{\partial \textbf{Q}}\cdot\frac{\textbf{k}-\textbf{Q}}{\varepsilon_{\textbf{k}}}\Bigg|_{\textbf{k}=\textbf{Q}}\nonumber\\
                 &=&\frac{A_{\textbf{Q}}}{S^{2}}\textbf{T}_{sc}(\textbf{Q})\cdot\frac{\textbf{k}-\textbf{Q}}{\varepsilon_{\textbf{k}}}\Bigg|_{\textbf{k}=\textbf{Q}}
\end{eqnarray}
The explicit expression along $k_{y}=0$ is
\begin{eqnarray}
  F(Q)&=&\textbf{sgn}(k_{x}-Q)\frac{A_{\textbf{Q}}}{v_{Q}}\sum_{\textbf{p}}\frac{A_{\textbf{p}}+B_{\textbf{p}}}{\varepsilon_{\textbf{p}}}\frac{\partial J_{\textbf{p}+\textbf{Q}}}{\partial Q} \nonumber\\
      &=&\textbf{sgn}(k_{x}-Q)\frac{2}{S}\frac{A_{\textbf{Q}}}{v_{Q}}T_{sc}(Q)
\end{eqnarray}
Here $\textbf{sgn}(x)$ represents the sign function and $v_{Q}$ is the spin-wave velocity at $k_{x}=Q$ along $k_{y}=0$.
The analytic result shows that our spin-wave spectrum behave like a step across $\textbf{k}=\textbf{Q}$ rather than a cone, which is consistent with our numerical result. More seriously, $F(Q)$ also blows up as $\alpha$ approaches the classical critical point $\alpha=2$ as shown in Fig.4.
\begin{figure}
  \centering\includegraphics[width=8.5cm]{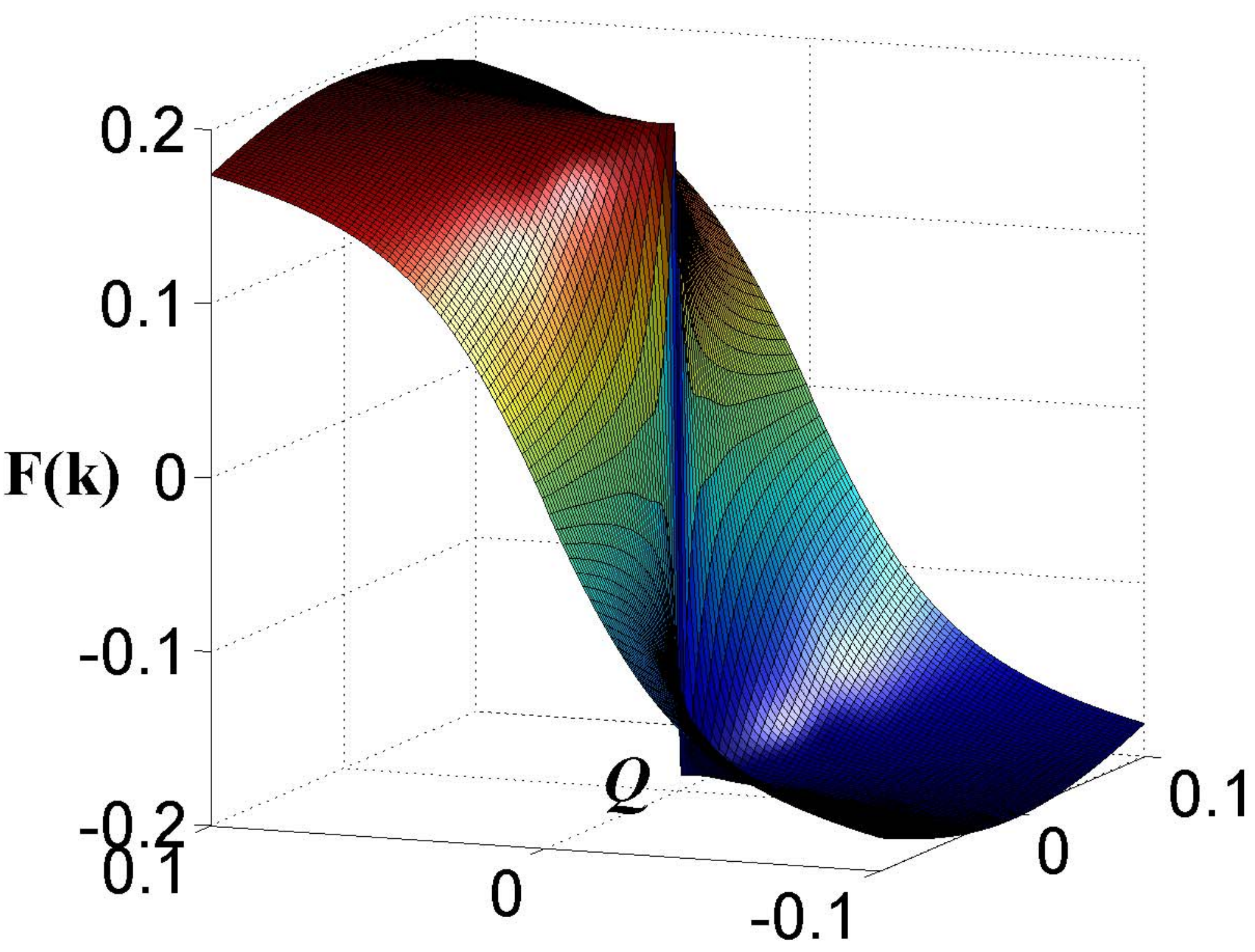}\\
  \caption{Numerical results of the spin-wave spectrum around $k=Q$ for $\alpha=0.5$.}
\end{figure}

Interestingly, this singular behavior is directly connected with the spin Casimir torque defined previously. It is indeed due to the ignorance of the spin Casimir effect in the above treatment. Since the spin Casimir torque is directly connected with the $1/S$ ordering vector correction $Q_{1}$, one suggests that the $Q_{1}$ induced modification of the linear spin-wave spectrum $\varepsilon_{k}$ may avoid the above problem in some region of the parameter space where $Q_{1}$ is still finite and reasonable.
In fact, with the equality
\begin{equation}
   v_{Q}=2\sqrt{\rho_{s}A_{\textbf{Q}}}
\end{equation}
the relationship between $F(Q)$ and $Q_{1}$ is apparently shown as:
\begin{equation}
   F(Q)=-\frac{sgn(k_{x}-Q)}{2S}v_{Q}\cdot Q_{1}
\end{equation}
On the other hand, a direct $1/S$ order correction due to the ordering vector modification is
\begin{equation}
   \delta F(Q)=\frac{\partial\varepsilon_{\textbf{k}}}{\partial Q}\Bigg|_{\textbf{k}=\textbf{Q}}\cdot Q_{1}
\end{equation}
which does not match our expectation.
In other words, this singular behavior can't be regularized in an conventional $1/S$ manner, and it can even lead to divergence in high order expansions.

\subsection{Sublattice Magnetization}
In this section, we turn to the sublattice magnetization which is the order parameter for general long-range ordered spin states. It can be used to probe the possible existence of quantum disordered phases. Within the spin-wave approach, its definition is
\begin{equation}
\langle S\rangle=S-\langle a^{\dagger}_{i}a_{i}\rangle=S-\delta S_{1}-\delta S_{2}
\end{equation}
Here the first quantum correction $\delta S_{1}$ is the linear spin-wave result given by
\begin{equation}
\delta S_{1}=\sum_{\textbf{k}}n_{\textbf{k}}
\end{equation}
And the second term is the second-order correction to the sublattice magnetization, which needs an evaluation of the one-loop results of the normal and anomalous Green's functions.~\onlinecite{Chubukov2,Sasha1} This term have three contributions
\begin{equation}
\delta S_{2}=\delta S^{1}_{2}+\delta S^{2}_{2}+\delta S^{3}_{2}
\end{equation}
These contributions are related to the normal and anomalous self-energies as shown in Fig.3(c), 3(d) and Fig.5.
The contributions from the normal self-energies are collected as $\delta S^{1}_{2}$, and those from the anomalous self-energies $O_{\textbf{k}}$ and $\Sigma^{c,d}_{3}$ are collected as $\delta S^{2}_{2}$, $\delta S^{3}_{2}$ respectively. Calculations of each contribution to the second-order correction are straightforward and the final results are
\begin{equation}
\delta S_{2}=\frac{1}{2S}\sum_{\textbf{k}}I_{tot}(\textbf{k}),~~~~~\delta S^{i}_{2}=\frac{1}{2S}\sum_{\textbf{k}}I_{i}(\textbf{k})
\end{equation}
with $i\in(1,2,3)$, $I_{tot}(\textbf{k})=\sum_{i}I_{i}(\textbf{k})$ and
\begin{eqnarray}
I_{1}(\textbf{k})&=&\frac{1}{2}\sum_{\textbf{p}}\frac{A_{\textbf{k}}}{\varepsilon_{\textbf{k}}}\cdot
                      \frac{|\Gamma_{2}(\textbf{k},\textbf{p})|^{2}}{(\varepsilon_{\textbf{k}}+\varepsilon_{\textbf{p}}+\varepsilon_{\textbf{k}+\textbf{p}})^{2}}\nonumber\\
I_{2}(\textbf{k})&=&\frac{1}{2}\sum_{\textbf{p}}\frac{B_{\textbf{k}}}{\varepsilon_{\textbf{k}}}\cdot
                      \frac{O_{\textbf{k}}}{\varepsilon^{2}_{\textbf{k}}} \nonumber\\
I_{3}(\textbf{k})&=&\frac{1}{2}\sum_{\textbf{p}}\frac{B_{\textbf{k}}}{\varepsilon_{\textbf{k}}}\cdot
                      \frac{\Gamma_{1}(\textbf{k},\textbf{p})\Gamma_{2}(-\textbf{k},\textbf{p})}{\varepsilon_{\textbf{k}}(\varepsilon_{\textbf{k}}+\varepsilon_{\textbf{p}}+\varepsilon_{\textbf{k}-\textbf{p}})}
\end{eqnarray}

As often taken for the high-order perturbative expansion, the integrands of all the three contributions are divergent.
Especially, the integrands $I_{2}(\textbf{k})$ and $I_{3}(\textbf{k})$  behave
as $O(1/k^{3})$ at $\textbf{k}\rightarrow\textbf{Q}$, implying that not only the leading
divergences in them, but also the subleading ones $O(1/k^{2})$
must cancel in order to produce finite result. Expanding near $\textbf{k}=\textbf{0}$ and $\textbf{k}=\textbf{Q}$ points, such a cancelation can be verified
analytically at $\alpha=1$ as shown by Chubukov.~\onlinecite{Chubukov2}
Here we perform the expansion with a general $\alpha$.

Because the divergence appears only near $\textbf{k}=\textbf{0}$ and $\textbf{k}=\textbf{Q}$, the terms proportional to $\varepsilon_{\textbf{k}}$ (finite part) can be ignored as before.
After lengthy but straightforward derivation one obtains
\begin{eqnarray}
I_{1}(\textbf{k})&\approx&\frac{A_{\textbf{k}}}{8\varepsilon_{\textbf{k}}}\Big[\kappa^{2}_{\textbf{k}}U_{0}(\textbf{k})+\gamma^{2}_{\textbf{k}}U_{Q}(\textbf{k})\Big] \nonumber\\
I_{2}(\textbf{k})&\approx&\frac{B_{\textbf{k}}}{4\varepsilon^{3}_{\textbf{k}}}\Big[-\kappa^{2}_{\textbf{k}}G_{0}(\textbf{k})+\gamma^{2}_{\textbf{k}}G_{Q}(\textbf{k})\Big] \nonumber\\
I_{3}(\textbf{k})&\approx&\frac{B_{\textbf{k}}}{8\varepsilon^{2}_{\textbf{k}}}\Big[\kappa^{2}_{\textbf{k}}V_{0}(\textbf{k})-\gamma^{2}_{\textbf{k}}V_{Q}(\textbf{k})\Big]
\end{eqnarray}
with
\begin{eqnarray}
   U_{0}(\textbf{k})&=&\sum_{\textbf{p}}\frac{\Gamma_{0}^{2}}{(\varepsilon_{\textbf{k}}+\varepsilon_{\textbf{p}}+\varepsilon_{\textbf{k}-\textbf{p}})^{2}}\nonumber\\
   U_{Q}(\textbf{k})&=&\sum_{\textbf{p}}\frac{\Gamma_{Q}^{2}}{(\varepsilon_{\textbf{k}}+\varepsilon_{\textbf{p}}+\varepsilon_{\textbf{k}-\textbf{p}})^{2}}   \nonumber\\
   V_{0}(\textbf{k})&=&\sum_{\textbf{p}}\frac{\Gamma_{0}^{2}}{\varepsilon_{\textbf{k}}+\varepsilon_{\textbf{p}}+\varepsilon_{\textbf{k}-\textbf{p}}}\nonumber\\
   V_{Q}(\textbf{k})&=&\sum_{\textbf{p}}\frac{\Gamma_{Q}^{2}}{\varepsilon_{\textbf{k}}+\varepsilon_{\textbf{p}}+\varepsilon_{\textbf{k}-\textbf{p}}}
\end{eqnarray}
where $G_{0}(\textbf{k})$, $G_{Q}(\textbf{k})$, $\Gamma_{0}$, and $\Gamma_{Q}$ have been defined in Eq.(29) and Eq.(30).
These equations are similar to those obtained in the expansion of spin-wave spectrum. As a consequence, the divergence cancelation results are quite similar as well. The leading and subleading parts of the integrands are zero at $\textbf{k}=\textbf{0}$ point, as expected.
However, the perfect divergent cancelation near $\textbf{k}=\textbf{Q}$ is not accessed in our general case and the final result is a subleading divergent contribution
\begin{eqnarray}
I^{div}_{tot}(\textbf{Q})&=&\frac{A_{\textbf{Q}}B_{\textbf{Q}}}{2\varepsilon^{2}_{\textbf{Q}}}
              \sum_{p}\frac{A_{\textbf{p}}+B_{\textbf{p}}}{\varepsilon_{\textbf{p}}}\frac{\partial J_{\textbf{p}+\textbf{Q}}}{\partial \textbf{Q}}
              \cdot\frac{\textbf{k}-\textbf{Q}}{\varepsilon_{\textbf{k}}}\Bigg|_{\textbf{k}=\textbf{Q}} \nonumber\\
              &=&\frac{A_{\textbf{Q}}B_{\textbf{Q}}}{S\varepsilon^{2}_{\textbf{Q}}}\textbf{T}_{sc}(\textbf{Q})\cdot\frac{\textbf{k}-\textbf{Q}}{\varepsilon_{\textbf{k}}}\Bigg|_{\textbf{k}=\textbf{Q}}
\end{eqnarray}
whose explicit expression along $k_{y}=0$ is
\begin{eqnarray}
I^{div}_{tot}(Q)&=&\textbf{sgn}(k_{x}-Q)\frac{B_{\textbf{Q}}}{\varepsilon^{2}_{\textbf{Q}}}\frac{SA_{\textbf{Q}}}{v_{Q}}\cdot
              \sum_{p}\frac{A_{\textbf{p}}+B_{\textbf{p}}}{\varepsilon_{\textbf{p}}}\frac{\partial J_{\textbf{p}+\textbf{Q}}}{\partial \textbf{Q}} \nonumber\\
              &=&\textbf{sgn}(k_{x}-Q)\frac{B_{\textbf{Q}}}{\varepsilon^{2}_{\textbf{Q}}}\frac{2A_{\textbf{Q}}}{v_{Q}}\cdot T_{sc}(Q) \nonumber\\
              &=&S\frac{B_{\textbf{Q}}}{\varepsilon^{2}_{\textbf{Q}}}\cdot F(Q)
\end{eqnarray}
which is related to the spin Caisimir torque too. One may argue that this divergent result is inherited from the singular behavior of the spin-wave spectrum. As a matter of fact, this divergence is indeed caused by the one-loop Green's function, but it is generated from the abnormal Green's function rather than the normal Green's function which accounts for the spin-wave spectrum.

Correspondingly, a conventional $1/S$ consideration of the ordering vector modification induced sublattice magnetization correction is ~\onlinecite{Da}
\begin{equation}
\delta S^{Q}_{2}=-\textbf{Q}_{1}\sum_{k}\frac{B_{\textbf{k}}(A_{\textbf{k}}+B_{\textbf{k}})}{4\varepsilon^{3}_{\textbf{k}}}\cdot\frac{\partial J_{\textbf{k}+\textbf{Q}}}{\partial \textbf{Q}}\Bigg|_{\textbf{Q}_{cl}}
\end{equation}
which is a finite integration. The total sublattice magnetization is still divergent even by adding this contribution, preventing the spin-wave expansion beyond the harmonic approximation in spiral phases. As we have shown, the only region in the parameter space that is free of divergence is the $\alpha=1$ point, but it is not likely that the long-range order only exists at this single point.~\onlinecite{Zheng1,Zheng2,VMC,FRG,MSW1,MSW2} Furthermore, one can expect more serious divergence when the expansion to higher order is carried out, making the conventional spin-wave theory failed. Accordingly, an alternative expansion scheme that can encompass the spin Casimir effect is needed.

\section{Torque Equilibrium Spin Wave Theory}
The spin Casimir effect described in our work can be generalized as the effect to shift the quantum fluctuation induced classical saddle point in quantum spin systems.
As a consequence, the corresponding spin excitations should be considered by expansion around the shifted (new) saddle point.
However, a shift of the saddle point is prevented by two essential issues:
1. Where is the new saddle point while the $1/S$ expansion results are divergent at some parameter region?
2. How to shift the saddle point from $\textbf{Q}_{cl}$ to $\textbf{Q}$ in the effective Hamiltonian with so many functions defined only at $\textbf{Q}_{cl}$?

The first issue can be easily handled using our torque description of the spin Casimir effect. The quantum fluctuation induced shift of the saddle point is accomplished by the spin Casimir torque. As this spin Casimir torque shifts the spin structure away from its classical saddle point, a classical reaction spin torque is generated by the deformation. The final saddle point is determined by the torque equilibrium condition
\begin{equation}
\textbf{T}_{sc}(\textbf{Q})+\textbf{T}_{cl}(\textbf{Q})=0
\end{equation}
where
\begin{equation}
\textbf{T}_{cl}(\textbf{Q})=\frac{\partial E_{cl}(\textbf{Q})}{\partial \textbf{Q}}=S^{2}\frac{\partial J_{\textbf{Q}}}{\partial \textbf{Q}}
\end{equation}
This is nothing but the variational equation of the vacuum energy $\delta E_{vac}/\delta \textbf{Q}=0$.
However, this equation can not be solved directly owing to the existence of the second issue, which is the main difficulty in our perturbative expansion scheme. To overcome it, an alternative spin-wave expansion scheme is needed.

It is noted that $A_{\textbf{k}}$ and $B_{\textbf{k}}$ are well-defined for all $\textbf{Q}$. The ordering vector can be shifted arbitrarily for the Hamiltonian $\hat{\mathcal{H}}_{2}$, $\hat{\mathcal{H}}_{3}$ and $\hat{\mathcal{H}}_{4}$ in Eq.(5).
Yet, for $\textbf{Q}\neq \textbf{Q}_{cl}$ the resultant $\varepsilon_{\textbf{k}}(\textbf{Q})$ would be imaginary at some $\textbf{k}$ points reminding that the expansion is carried out around the wrong saddle point.
The basic idea of our scheme is to rearrange the perturbation processes and take into account the ordering vector modification in an self-consistent way.

Given $\textbf{Q}$ as the final ordering vector of the system, all the functions in Eq.(5) are shifted from $\textbf{Q}_{cl}$ to $\textbf{Q}$. For any function, taking $A_{\textbf{k}}$ as an example here, it can be written as
\begin{equation}
A_{\textbf{k}}(\textbf{Q})=\widetilde{A}_{\textbf{k}}(\textbf{Q})+A^{c}_{\textbf{k}}(\textbf{Q})
\end{equation}
with
\begin{equation}
A^{c}_{\textbf{k}}(\textbf{Q})=A_{\textbf{k}}(\textbf{Q})-\widetilde{A}_{\textbf{k}}(\textbf{Q})
\end{equation}
where $\widetilde{A}_{\textbf{k}}$ is the $A_{\textbf{k}}$ function of another spin system whose classical ordering vector $\widetilde{\textbf{Q}}_{cl}$ equals to $\textbf{Q}$.
Here we assume that this spin system has the same symmetry and set of exchange integrals as the original one. As a result, this spin system is nothing but the original one with a different $\alpha$ denoted as $\widetilde{\alpha}$.
The spin-wave Hamiltonian written in this manner is
\begin{equation}
\hat{\mathcal{H}}_{2}(\alpha,\textbf{Q})=\widetilde{\mathcal{H}}_{2}(\widetilde{\alpha},\textbf{Q})+\mathcal{H}^{c}_{2}
\end{equation}
with
\begin{eqnarray}
\widetilde{\mathcal{H}}_{2}&=&2S\sum\limits_{\textbf{k}}\widetilde{A}_{\textbf{k}}a^{\dagger}_{\textbf{k}}a_{\textbf{k}}-\frac{\widetilde{B}_{\textbf{k}}}{2}(a_{\textbf{k}}a_{-\textbf{k}}+a^{\dagger}_{\textbf{k}}a^{\dagger}_{-\textbf{k}})
\nonumber\\
\mathcal{H}^{c}_{2}&=&2S\sum\limits_{\textbf{k}}A^{c}_{\textbf{k}}a^{\dagger}_{\textbf{k}}a_{\textbf{k}}-\frac{B^{c}_{\textbf{k}}}{2}(a_{\textbf{k}}a_{-\textbf{k}}+a^{\dagger}_{\textbf{k}}a^{\dagger}_{-\textbf{k}})
\end{eqnarray}
The $\widetilde{\mathcal{H}}_{2}$ is free of imaginary energy problem and $\mathcal{H}^{c}_{2}$ is obviously proportional to $\textbf{T}_{sc}$.
In spite of the $2S$ factor, the Hamiltonian $\mathcal{H}^{c}_{2}$ is of order $O(S^{0})$ just as $\hat{\mathcal{H}}_{4}$ due to the fact that $A^{c}_{\textbf{k}}(\textbf{Q})$ and $B^{c}_{\textbf{k}}(\textbf{Q})$ are of order $O(1/S)$.
In the spirit of the $1/S$ expansion, we treat $\mathcal{H}^{c}_{2}$ as an interaction term and $\widetilde{\mathcal{H}}_{2}$ as the modified harmonic Hamiltonian. The spin Casimir torque at the equilibrium point can be expressed approximately as
\begin{equation}
   \widetilde{\textbf{T}}_{sc}(\textbf{Q})
            =\frac{S}{2}\sum\limits_{\textbf{k}}\frac{\widetilde{A}_{\textbf{k}}+\widetilde{B}_{\textbf{k}}}{\widetilde{\varepsilon}_{\textbf{k}}}\cdot\frac{\partial \widetilde{J}_{\textbf{k}+\textbf{Q}}}{\partial \textbf{Q}}
\end{equation}
This approximate spin Casimir torque has only one component $\widetilde{T}_{sc}(Q)$ as well.
And the torque equilibrium equation can be approximately written as
\begin{equation}
\frac{\partial J_{\textbf{Q}}}{\partial \textbf{Q}}=-\frac{1}{2S}\sum\limits_{\textbf{k}}\frac{\widetilde{A}_{\textbf{k}}+\widetilde{B}_{\textbf{k}}}{\widetilde{\varepsilon}_{\textbf{k}}}\cdot\frac{\partial \widetilde{J}_{\textbf{k}+\textbf{Q}}}{\partial \textbf{Q}}
\end{equation}
The exchange parameters on the left hand side of the equation is exact as $\alpha$ while the parameters on the right hand side approximate as $\widetilde{\alpha}$.
Noticing that $\widetilde{\alpha}=-2\cos(Q/2)$, this equation can be solved and the numerical results are shown in Fig.(2).
The results show drastic modification of the classical ordering vector caused by the quantum fluctuation.
And our results are similar with the results obtained by much more sophisticated numerical methods.~\onlinecite{Zheng1,MSW1,MSW2}
Interestingly, for $\alpha>1.2$, the torque equilibrium equation has no non-trivial solutions other that $Q=2\pi$, hence no spiral order is stable.
Therefore, in this region, the system is either in a QObD induced Neel phase or some quantum disordered phases.
Nevertheless, whether a spin liquid phase exists in this region is still under controversial.~\onlinecite{Zheng1,FRG,MSW1,MSW2}
Either way, the conventional spin-wave expansion starting from the classical spiral state is qualitatively incorrect for $1.2<\alpha<2$.
As $S$ becomes large, the range of this spiral instable region is narrowed and eventually shrinks to a point $\alpha=2$ for $S=\infty$.

\begin{figure}
  \includegraphics[width=8.5cm]{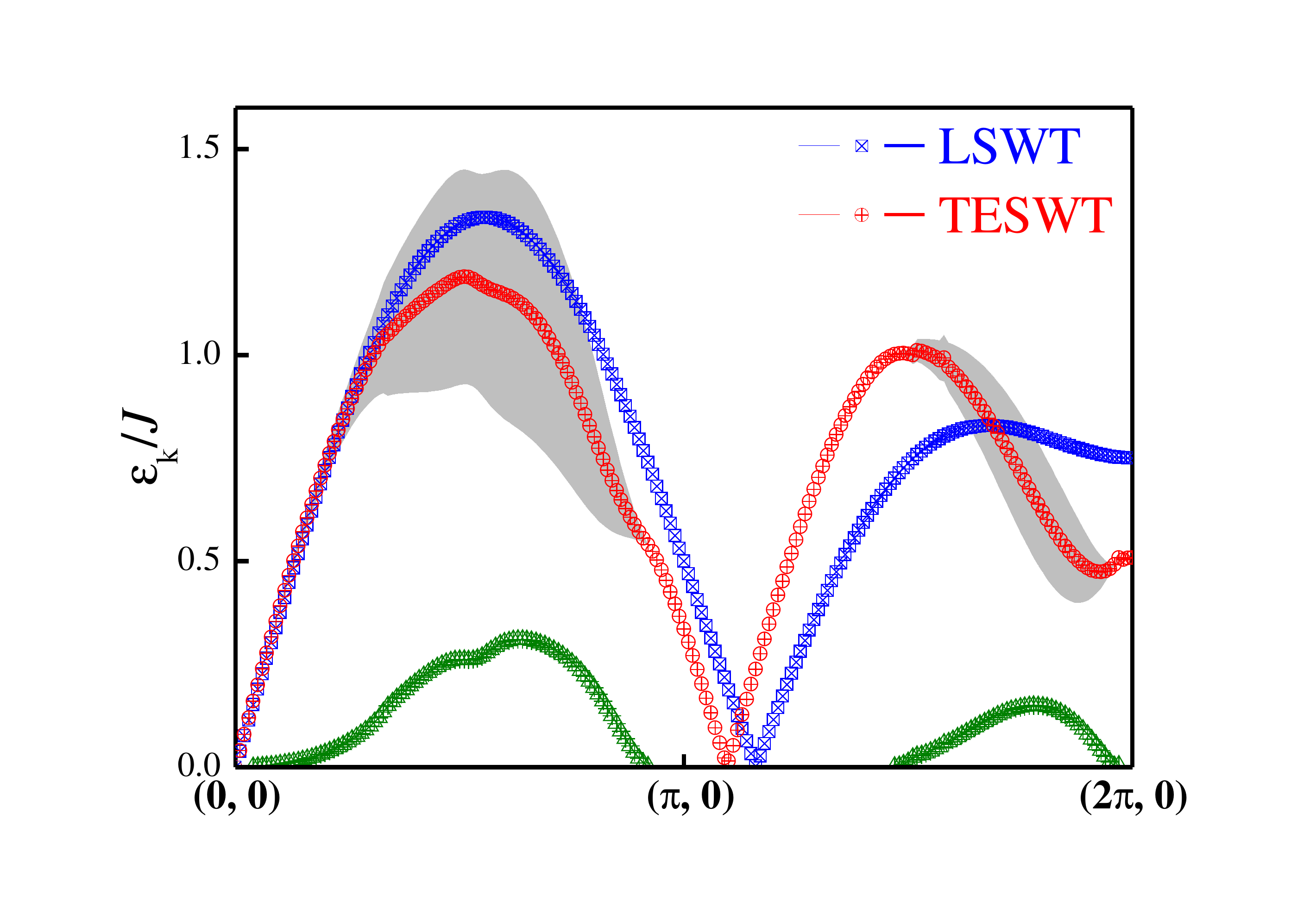}\\
  \caption{Spin-wave spectrum for $\alpha=0.5$ along the symmetric directions in the Brillouin zone. The blue line is the results obtained in the LSWT approximation. The red and green lines are the real (energy) and imaginary (damping rate) parts of our $1/S$ results, respectively. And the gray areas show the width of the spectral peaks due to the damping.~\onlinecite{Sasha1}}
\end{figure}

The above separation procedure can be repeated to every term in Eq.(5). Eventually, we obtain a whole set of $\widetilde{\mathcal{H}}_{i}$ and $\mathcal{H}^{c}_{i}$. The resultant new terms are very similar to the counterterms in quantum field theory.~\onlinecite{QFT}
Interestingly, the Casimir problem in the quantum field theory is divergent and this divergence can be regulated by introducing the counter-terms that are fixed with physical renormalization conditions.~\onlinecite{CaFT1,CaFT2} It is expected that our divergence problem can be solved in a similar way.
Given that we are only interested in the results at $1/S$ order, the "counter-terms" $\mathcal{H}^{c}_{3}$ and $\mathcal{H}^{c}_{4}$ can be neglected.
\begin{equation}
    \widetilde{\mathcal{H}}_{sw}=\widetilde{\mathcal{H}}_{2}+\mathcal{H}^{c}_{2}+\widetilde{\mathcal{H}}_{3}+\widetilde{\mathcal{H}}_{4}
\end{equation}
Following the same procedure described in Section III, the effective Hamiltonian reads
\begin{eqnarray}
   \widetilde{\mathcal{H}}_{eff}&=&\sum\limits_{\textbf{k}}\Bigg\{(2S\widetilde{\varepsilon}_{\textbf{k}}+\delta\widetilde{\varepsilon}_{\textbf{k}})b^{\dagger}_{\textbf{k}}b_{\textbf{k}}
                   -\frac{\widetilde{O}_{\textbf{k}}}{2}(b_{\textbf{k}}b_{-\textbf{k}}+b^{\dagger}_{\textbf{k}}b^{\dagger}_{-\textbf{k}})\nonumber\\
                &&+2S\Big[\varepsilon^{c}_{\textbf{k}}b^{\dagger}_{\textbf{k}}b_{\textbf{k}}
                   -\frac{O^{c}_{\textbf{k}}}{2}(b_{\textbf{k}}b_{-\textbf{k}}+b^{\dagger}_{\textbf{k}}b^{\dagger}_{-\textbf{k}})\Big]\Bigg\}\nonumber\\
                   \nonumber\\
                &&+i\sqrt{2S}\sum\limits_{\textbf{k},\textbf{p}}\Big[\frac{1}{2!}\widetilde{\Gamma}_{1}(\textbf{p},\textbf{k}-\textbf{p};\textbf{k})b_{\textbf{k}}b^{\dagger}_{\textbf{k}-\textbf{p}}b^{\dagger}_{\textbf{p}} \nonumber\\
               &+&\frac{1}{3!}\widetilde{\Gamma}_{2}(\textbf{p},-\textbf{k}-\textbf{p};\textbf{k})b^{\dagger}_{\textbf{p}}b^{\dagger}_{-\textbf{k}-\textbf{p}}b^{\dagger}_{\textbf{k}}-\textrm{H.c.}\Big]
\end{eqnarray}
Here $\widetilde{R}$ represents $R(\widetilde{\alpha},\textbf{Q})$ and
\begin{eqnarray}
    \varepsilon^{c}_{\textbf{k}}&=&(\widetilde{u}^{2}_{\textbf{k}}+\widetilde{v}^{2}_{\textbf{k}})A^{c}_{\textbf{k}}-2\widetilde{u}_{\textbf{k}}\widetilde{v}_{\textbf{k}}B^{c}_{\textbf{k}} \nonumber\\
    O^{c}_{\textbf{k}}&=&(\widetilde{u}^{2}_{\textbf{k}}+\widetilde{v}^{2}_{\textbf{k}})B^{c}_{\textbf{k}}-2\widetilde{u}_{\textbf{k}}\widetilde{v}_{\textbf{k}}A^{c}_{\textbf{k}}
\end{eqnarray}
The diagram representation of these extra terms are similar to the counter-terms in the quantum field theory as shown in Fig.3(e) and 3(f). The resultant contribution from our "counter-terms" to the $1/S$ spin-wave spectrum is
\begin{equation}
   \varepsilon^{c}_{\textbf{k}}=\frac{1}{\widetilde{\varepsilon}_{\textbf{k}}}\Big[\widetilde{A}_{\textbf{k}}A_{\textbf{k}}(\textbf{Q})-\widetilde{B}_{\textbf{k}}B_{\textbf{k}}(\textbf{Q})\Big]-\widetilde{\varepsilon}_{\textbf{k}}
\end{equation}
whose contribution to $\textbf{k}=\textbf{0}$ point is zero while that to $\textbf{k}=\textbf{Q}$ point is
\begin{equation}
   F^{c}(\textbf{Q})=\widetilde{A}_{\textbf{Q}}\frac{\partial J_{\textbf{Q}}}{\partial \textbf{Q}}\cdot\frac{\textbf{k}-\textbf{Q}}{\widetilde{\varepsilon}_{\textbf{k}}}\Bigg|_{\textbf{k}=\textbf{Q}}
\end{equation}
And the contribution from the other terms is
\begin{equation}
  \widetilde{F}(\textbf{Q})=\frac{\widetilde{A}_{\textbf{Q}}}{S^{2}}\widetilde{\textbf{T}}_{sc}(\textbf{Q})\cdot\frac{\textbf{k}-\textbf{Q}}{\widetilde{\varepsilon}_{\textbf{k}}}\Bigg|_{\textbf{k}=\textbf{Q}}
\end{equation}
Noting that the approximate torque equilibrium equation Eq.(56) can be written as $S^{2}\partial J_{\textbf{Q}}/\partial \textbf{Q}=-\widetilde{\textbf{T}}_{sc}(\textbf{Q})$, the singular contribution is canceled by the "counter-terms" contribution.
The cancelation along other directions is straightforward to certificate and the Goldstone theorem is preserved as it should be.
The numerical results of the spin-wave spectrum are shown in Fig.7, which manifests well behaved Goldstone modes and magnon decay effects.~\onlinecite{Sasha1,Sasha2,Sasha3,Oleg}
The downward modification of the magnon spectrum and the line shape of the magnon damping rate are consistent with previous works~\onlinecite{Sasha1,Sasha2}.
For $\alpha=1$, the spin Casimir effect vanishes and our results are exactly the same with that obtained by Chernyshev and Zhitomirsky.~\onlinecite{Sasha1,Sasha2,Sasha3}

On the other hand, the contribution to the sublattice magnetization from our "counter-terms" is
\begin{equation}
\delta S^{c}_{2}=\frac{1}{2S}\sum_{\textbf{k}}I_{c}(\textbf{k})
\end{equation}
with
\begin{equation}
I_{c}(\textbf{k})=\frac{1}{2}\frac{B_{\textbf{k}}}{\varepsilon_{\textbf{k}}}\cdot\frac{O^{c}_{\textbf{k}}}{\varepsilon^{2}_{\textbf{k}}}
\end{equation}
This integrand is also zero at $\textbf{k}=\textbf{0}$ point and divergent at $\textbf{k}=\textbf{Q}$ point as
 \begin{equation}
I_{c}(\textbf{Q})=S\frac{\widetilde{A}_{\textbf{Q}}\widetilde{B}_{\textbf{Q}}}{
          \widetilde{\varepsilon}^{2}_{\textbf{Q}}}\frac{\partial J_{\textbf{Q}}}{\partial \textbf{Q}}
          \cdot\frac{\textbf{k}-\textbf{Q}}{\widetilde{\varepsilon}_{\textbf{k}}}\Bigg|_{\textbf{k}=\textbf{Q}}
\end{equation}
And the divergent contribution from the other terms are
 \begin{equation}
\widetilde{I}^{div}_{tot}(\textbf{Q})=\frac{\widetilde{A}_{\textbf{Q}}\widetilde{B}_{\textbf{Q}}}{S\widetilde{\varepsilon}^{2}_{\textbf{Q}}}
                                       \widetilde{\textbf{T}}_{sc}(\textbf{Q})
                                       \cdot\frac{\textbf{k}-\textbf{Q}}{\widetilde{\varepsilon}_{\textbf{k}}}\Bigg|_{\textbf{k}=\textbf{Q}}
\end{equation}
These two divergences cancel again owing to the approximate torque equilibrium equation and this cancelation persists to other directions.
Eventually, we can obtain a finite second order correction to the sublattice magnetization.

\begin{figure}
  \includegraphics[width=8.5cm]{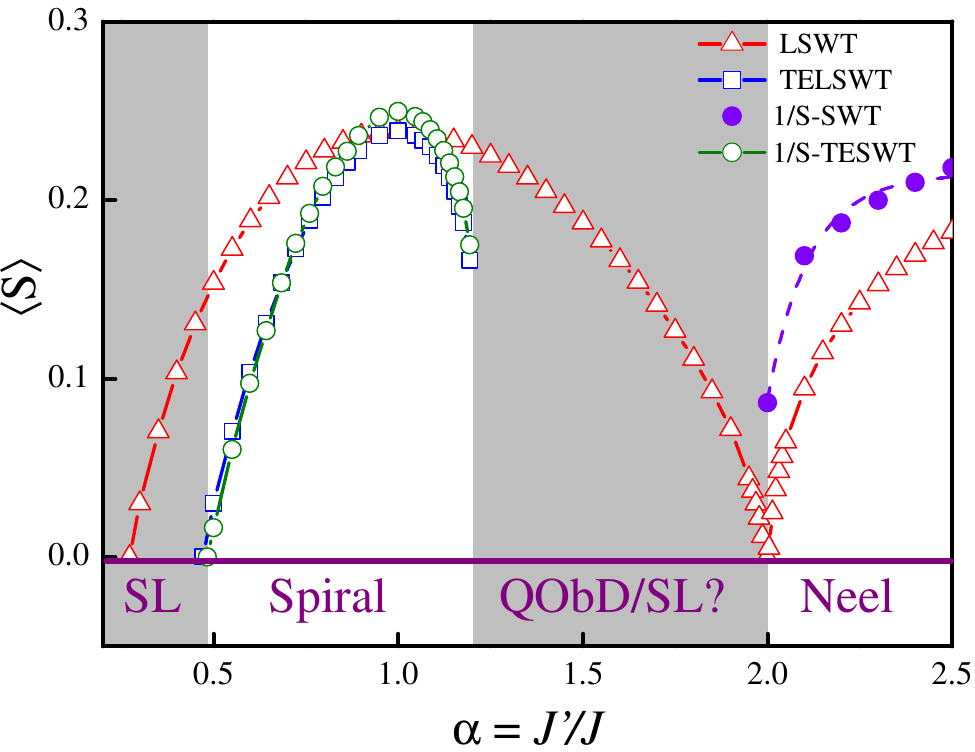}\\
  \caption{The sublattice magnetization $\langle S\rangle$ as a function of $\alpha=J^{\prime}/J$.
The red triangular dots represent the LSWT results, and the purple red dots represent the standard $1/S$ expansion results.
The blue square dots are our torque equilibrium linear spin-wave theory (TELSWT) results, and the green red dots are our $1/S$-TESWT expansion results.}
\end{figure}

The above results allow us to perform systematic calculations on the sublattice magnetization of the system in the whole parameter space, which may indicate potential existence of spin liquid phases.~\onlinecite{QM}
The calculation is straightforward and follows our previous established scheme in the spiral region and the sublattice magnetization in the Neel state is obtained within a standard spin-wave expansion scheme. The final results of the sublattice magnetization and related phase diagram are presented in Fig.8.
In the obtained quantum phase diagram, the spiral state is destroyed by quantum fluctuation as $\alpha<0.5$, which is not far from the MSW results. In addition, above the isotropic point $\alpha=1$, the spiral state is no longer stable in the region $\alpha>1.2$, which is also consistent with previous numerical studies.~\onlinecite{Zheng1,MSW1} This instability is not due to the vanishing sublattice magnetization but owing to the fact that the classical saddle surface is so shallow that quantum fluctuation can induce the instability of the classical spiral state.
In our torque description, this instability can be understood as the so weak stiffness of the "spring" that the spin Casimir torque can squeeze it arbitrarily until the torque disappears. In this manner, the spin Casimir effect shows us a spatial way to "melt" a long-range non-collinear ordered state, which is different with the perception of the collinear cases.
What should be mentioned here is that the exact structure of the phase diagram in the region $1.2<\alpha<2$ is still unclear,~\onlinecite{Zheng1,FRG,MSW1} and the perturbative nature of our approach prevents us to identify whether the system is in a QObD induced Neel phase or some quantum disordered phases.

Our approach can be easily generalized to systems with multi parameters by considering more counter-terms such as $\mathcal{H}^{c}_{3}$ and $\mathcal{H}^{c}_{4}$. Other than that, more self-consistent equations can be obtained using the divergence cancelation condition in higher order. The non-renormalizable form of the effective Hamiltonian ensures that we can have all kinds of divergences and counter-terms as long as we carry out our calculations to high enough order.~\onlinecite{QFT} However, the whole spin-wave basis can breakdown if high order loops are considered and it is usually unnecessary.~\onlinecite{RMP1} One can simply apply the present scheme to the parameter dominating the ordering vector, given other parameters unchanged. In addition, our approach can be considered as a general approximation method for the quantum fluctuation induced saddle point shift problems.

\section{Experimental Applications}
Due to the fundamental theoretical interest, further impetus to investigate frustrated magnets has arisen from recent experimental developments identifying several materials in which spin liquid like unconventional behaviors is observed.~\onlinecite{TL1,TL2}
Hence potential progress is expected in the near future. So as to test the theory experimentally, a precise information on the spin-Hamiltonian parameters for the materials of interest is demanded. And the most effective way of solving this problem is to suppress quantum fluctuation by strong-enough magnetic field. However, such experiments are only possible for those systems with small-enough exchange parameters such as $Cs_{2}CuCl_{4}$, so that the required field can be available practically. It would be highly desirable to dispose of a fast tool that can estimate the correct exchange parameters and outline the quantum phase diagrams.

Here we propose that the methods presented in this work will serve this very purpose.
Compared with other sophisticated numerical and analytical methods, our approach is much less technique involved.
As a matter of fact, detailed results can be obtained in our approach by calculation that is no harder than a linear spin-wave expansion.
Other than that, our approach can be conveniently applied to magnetic systems with general spin Hamiltonian and the only requirement is the existence of the long-range order.
In other words, our scheme only requires experimental results with magnetic fields that can drive the system into long-range ordered states.
More importantly, the long-range ordered phases in magnetic fields are usually non-collinear, such as "fan" and "umbrella" phases.
Add it all up, our method shows clinical improvement compared with the conventional parameter fitting processes based on LSWT but keeps the computational difficulty nearly unchanged.

In order to demonstrate the superiority of our method, we consider $Cs_{2}CuCl_{4}$ as an example.
The effective exchange parameters obtained from the global fit of the zero-field neutron scattering results are $\alpha=0.175$.~\onlinecite{CsCuCl3}
This fitting result based on LSWT is far from the exact result $\alpha=0.34$ that obtained using the high-field technique where quantum fluctuations are quenched out.~\onlinecite{CsCuCl1,CsCuCl2}
However, by using our approximate torque equilibrium equation Eq.(56), the LSWT fitting result $\widetilde{\alpha}=0.175$ predicts the bare parameters with $\alpha=0.35$, which is very close to the high-field exact result.
Beyond that, our approach further shows that the sublattice magnetization of the system without Dzyaloshinskii-Moriya (DM) interaction vanishes based on linear approximation.
As a result, the long-range spiral state in $Cs_{2}CuCl_{4}$ can't be stable without DM interaction, which is consistent with other theoretical analysis.~\onlinecite{Da,Vei}

In summary, our approach enables one to estimate the correct exchange parameters and outline the quantum phase diagrams based on simple linear spin-wave approximation.
Thus, besides the theoretical setups, our method can also serve as an efficient tool for the experimental fitting processes of the exchange parameters in general frustrated quantum magnets.

\section{Summary}
We have presented a detailed analysis of the spin-wave expansion on the spatial anisotropic triangular lattice Heisenberg antiferromagnets.~\onlinecite{TL1,TL2}
The phenomena that the classical ordering vector is modified by the quantum fluctuation is carefully studied and its Casimir nature is revealed. This effect shares the same origin with the well-known QObD effect.~\onlinecite{OBD1,OBD2}
Both of these cases can be interpreted as the spin Casimir effects in which the quasi-particle vacuum with zero-point fluctuation plays the role of the fluctuating vacuum and the classical spin structure act as the macroscopic boundary.
To describe these spin Casimir effects quantitatively we further define a spin Casimir torque, which describes a long-range torque effect generated by quantum spin fluctuation.

Base on these results, it is shown that the presence of the spin Casimir effect can induce divergent results in a conventional spin-wave expansion even though the long-range order is stable.
A careful expansion shows these divergences are directly connected with the spin Casimir torque.
As a result, the appearance of these divergences invalidate the conventional $1/S$ expansion in an obvious way.~\onlinecite{RMP1,QM}
And the encountered problems are rather generic and common to a variety of frustrated antiferromagnets regardless of the spin value.
For the systems with large spins, the situation is especially aggravating even though the spin Casimir torque goes to zero as $S\rightarrow\infty$.
This is due to the fact that long-range order is more stable and the $1/S$ expansion turns to be more reliable when $S\gg1$ while the divergence prevent any reasonable prediction.

In the present work, we have developed a self-consistent approach in the frame of the spin-wave theory, which is applicable to variety of systems with quantum fluctuation induced saddle point shift problems.
The self-consistently calculated modification to the ordering vector is finite and close to previous SE and MSW results as shown in Fig.(2).
Furthermore, our approach regularizes all the divergences in the $1/S$ expansion effectively.
This accomplishment allows us to calculate many spin-wave properties beyond linear approximation.
The spin-wave spectrum results are shown in Fig.7, which present well behaved Goldstone modes and mangnon decay effects.~\onlinecite{Sasha1,Sasha2,Sasha3,Oleg}
Other than the conventional spin-wave properties, an approximate quantum phase diagram is also obtained, which displays good consistency with previous numerical works.~\onlinecite{Zheng1,MSW1}
These results evidence that our approach is a suitable tool to study various other problem in non-collinear quantum antiferromagnets.
Besides the theoretical setups, our method can be further applied to estimate the correct exchange parameters and outline the quantum phase diagrams, which can be useful for experimental fitting processes in frustrated quantum magnets.

\section{Acknowledgment}

This work was supported by the Natural Science Foundation of China (Grant Nos. 11234005 and 51431006) and the National 973 Projects of China (Grant No. 2011CB922101).



\end{document}